\documentclass[fleqn,usenatbib]{mnras}

\usepackage{newtxtext,newtxmath}
\usepackage[T1]{fontenc}

\DeclareRobustCommand{\VAN}[3]{#2}
\let\VANthebibliography\thebibliography
\def\thebibliography{\DeclareRobustCommand{\VAN}[3]{##3}\VANthebibliography}

\usepackage{graphicx}	% Including figure files
\usepackage{amsmath}	% Advanced maths commands
\usepackage{mwe}
\usepackage{subcaption}
\usepackage{newtxtext,newtxmath}
\usepackage{threeparttable}
\usepackage{longtable}
\newcommand{\fref}[1]{Fig.~\ref{#1}}
\newcommand{\tref}[1]{Table~\ref{#1}}

\newcommand{\cref}[1]{Chapter~\ref{#1}}
\newcommand{\sref}[1]{Section~\ref{#1}}
\newcommand{\aref}[1]{Appendix~\ref{#1}}

\usepackage[dvipsnames]{xcolor}
\usepackage{xspace}

\newcommand{\comment}[1]
{\color{black}#1\color{black}\xspace}

\usepackage{bm}
\usepackage{enumitem}
\usepackage{stfloats}

\defcitealias{Qin2022}{Q22}

\title[SN Ia Siblings]{An archival search for type Ia supernova siblings}

\author[L. Kelsey]{
L.~Kelsey$^{1}$\thanks{E-mail: lisa.kelsey@port.ac.uk}
\\
$^{1}$ Institute of Cosmology and Gravitation, University of Portsmouth, Portsmouth, PO1 3FX, UK}

\date{Accepted XXX. Received YYY; in original form ZZZ}

\pubyear{2023}

\begin{document}
\label{firstpage}
\pagerange{\pageref{firstpage}--\pageref{lastpage}}
\maketitle

\begin{abstract}

By searching the Open Supernova Catalog, an extra-galactic transient host galaxy database, and literature analyses, I present the largest sample of type Ia supernova (SN Ia) siblings to date. The sample comprises 158 galaxies, consisting of 327 confirmed SNe Ia - over 10 times larger than existing sibling SN Ia samples. SN siblings share host galaxies, and thus share global environmental properties and associated systematic uncertainties. This makes them valuable for both cosmological and astrophysical analyses; for example, sibling SNe Ia allow for comparison of environmental properties within the same galaxy, progenitor comparisons, rates analyses, and multiple calibrations of the Hubble-Lema\^{\i}tre constant. This large sample will provide a variety of new avenues of research, and be of great interest to the wider SN Ia community. To give an example use of this sample, I define a cosmology sub-sample of 44 siblings; and use it to compare light-curve properties between sibling pairs. I find no evidence for correlations in stretch ($x_1$) and colour ($c$) between pairs of siblings. Moreover, by comparing to a comparable set of random pairs of SNe Ia through boot-strapping, I find that siblings are no more similar in $x_1$ and $c$ than any random pair of SNe Ia. Given that siblings share the same hosts, differences in $x_1$ and $c$ between siblings cannot be due to global galaxy properties. This raises important questions regarding environmental systematics for SN Ia standardisation in cosmology, and motivates future analyses of sibling SNe Ia.  

\end{abstract}


%%%%%%%%%%%%%%%%%%%%%%%%%%%%%%%%%%%%%%%%%%%%%%%%%%

%%%%%%%%%%%%%%%%% BODY OF PAPER %%%%%%%%%%%%%%%%%%

\section{Introduction} \label{intro}

As more supernovae are discovered in wide-field time domain surveys, the likelihood of finding multiple type Ia supernovae (SNe Ia) that have occurred in the same galaxy is increasing. This likelihood will only further increase with the dawn of the next generation of transient surveys, with the Rubin Observatory Legacy Survey of Space and Time (LSST) predicted to find $\sim800$ galaxies hosting multiple SNe Ia \citep{Scolnic2020}. These SNe are termed \lq{siblings}\rq\ \citep{Brown2014}, and they provide a novel data set to study the effects of galaxy environmental properties on the standardisation of SNe Ia in cosmology, and to provide further insight into potential progenitor scenarios. 

SNe Ia are important cosmological probes as standardisable candles due to their low intrinsic peak absolute magnitude dispersion. This dispersion can be further reduced using standard corrections based on correlations between SN Ia brightness and light-curve stretch \lq{$x_1$}\rq\ \citep[the \lq{brighter-slower}\rq\ relation;][]{Rust1974,Pskovskii1977,Phillips1993} and brightness and optical colour \lq{$c$}\rq\ \citep[the \lq{brighter-bluer}\rq\ relation;][]{Riess1996, Tripp1998}. 

Despite these corrections, there is a remaining brightness dispersion currently unaccounted for. A potential cause of this remaining dispersion is considered to be due to host galaxy effects because of clear correlations between brightness (or \lq{Hubble residual}\rq\ - the difference between observed and model brightness) and host galaxy properties. The most well-known of these is the \lq{mass step}\rq\, in which SNe Ia associated with high-mass host galaxies standardise to brighter luminosities than those in lower-mass hosts \citep{Sullivan2010,Kelly2010,Lampeitl2010,Gupta2011,Johansson2013,Childress2013,Uddin2017,Uddin2020,Smith2020,Ponder2020, Popovic2021}. Alongside this stellar mass correlation, other environmental properties such as star formation rate and galaxy colour also show this clear correlation \citep[e.g., ][]{Lampeitl2010,Sullivan2010,DAndrea2011,Childress2013,Pan2014,Wolf2016,Uddin2017,Kim2019,Kelsey2021,Kelsey2022}. In modern cosmological analyses, a host galaxy correction term is included in the cosmological fit, typically applying two different corrections for high- and low-mass hosts. 

As sets of SN Ia siblings share the same host galaxies, they share global environmental properties, along with systematic uncertainties from redshift, peculiar velocities and gravitational lensing. Therefore, any remaining differences in brightness, $x_1$, $c$, or Hubble residual between sets of siblings cannot be due to global host galaxy properties, potentially uncovering a concern with the current formalism in large surveys where siblings are corrected for in the same way. Instead, these differences may be due to sub-galactic differences in environments, dust distributions, progenitor scenarios, explosion mechanisms, or as-yet unknown physics. Thus, the use of siblings could determine if sub-galactic differences are the dominant factor in environmental standardisation, as suggested by local environment studies such as \citet{Rigault2013,Rigault2015,Rigault2020} and \citet{Kelsey2021,Kelsey2022}. 

By comparing eight sibling pairs from the Dark Energy Survey (DES), \citet{Scolnic2020} found no better agreement between SN properties (with the tentative exception of light-curve stretch, $x_1$) in the same galaxy as from any random pair of galaxies \citep[as was also found by][for a similar analysis using the Pantheon+ sample]{Scolnic2022}, and concluded that at least half of the intrinsic scatter in Hubble residuals is not from global host galaxy properties. \citet{Scolnic2020} also found that the sibling pair in the DES sample with the most differing distance modulii have SNe located on opposite sides of their host galaxy. Similarly, by finding a difference of 2 mag in rest-frame \textit{B}-band peak magnitude between another set of siblings, with one SN being very red (high $c$ value) whilst having a similar $x_1$ value to its sibling, \citet{Biswas2022} found evidence of sibling SNe experiencing differing levels of dust attenuation, as was also suggested by \citet{Elias-Rosa2008} for SN2002cv and SN2002bo in NGC 3190. 

\citet{Gall2018} found that distances for siblings SN2007on and SN2011iv in NGC 1404 differed by up to $14\%$, despite having similar decline rates. By studying SN1980N and SN1981D in NGC 1316, \citet{Hamuy1991} found that distances for the two siblings differed by $\sim 0.1$mag. When this was expanded by \citet{Stritzinger2010} to include the newer siblings in NGC 1316 (SN2006dd and SN2006mr), this distance dispersion was reduced to $\sim 4-8\%$. \citet{Hoogendam2022} found that the fast-declining SN siblings SN1997cn and SN2015bo (which also exhibited near-identical spectra) were consistent in distance at the $0.06\sigma$ level, and with distances derived using surface-brightness fluctuations, perhaps indicative that they exploded under similar circumstances. \comment{Additionally, \citet{DerKacy2022} measured distance modulii to NGC 5018 using the siblings SN2002dj and SN2021fxy and found them to be consistent at $1.2\sigma$}. Furthermore, \citet{Burns2020} found that a sample of 12 literature siblings gave an average consistency in distance of $3\%$. \citet{Ward2022} also found distances for a trio of SNe Ia (SN1997bq, SN2008fv, and SN2021hpr) in NGC 3147 were consistent, with a low standard deviation of $\leq 0.01$ mag. This range of findings across the literature gives clear motivation for the further analysis of larger samples of sibling SNe Ia.

Complementary to their potential to improve cosmological standardisation, SN Ia siblings are particularly useful for cosmology by providing multiple independent distance measurements to the same host galaxy if they also share their host with a Cepheid variable star. This gives a new, innovative calibration method for the Hubble-Lema\^{\i}tre constant ($H_0$, the rate of the expansion of the Universe). By calibrating the zero-point of Hubble diagrams from large surveys using individual sibling SN Ia independently, different values of $H_0$ are measured, providing uncertainties on the overall measurement \citep{Gallego-Cano2022}. With a large sample of siblings, this can be compared for each set, providing a new independent measurement of $H_0$ which may help to understand the remaining Hubble Tension. 

Alongside their use for cosmology, SN Ia siblings are valuable tools for investigating potential progenitor stellar populations by comparing differences in star formation rate, dust extinction and metallicity. Although not the focus of this work, by compiling samples of galaxies that contain many different types of SNe (i.e., not just SNe Ia) as has been done for the Zwicky Transient Facility (ZTF) \citep{Graham2022} or for a historical data set \citep{AndersonSoto2013}, understanding can be gained into rates of SNe \citep{Graur2017a,Graur2017b}, and preferences for different classes of SN to explode in different types of galaxies. 

However, despite the quality and wealth of data for these siblings found in recent large surveys such as DES and ZTF, they suffer from the same critical flaw. The rate of detection of siblings in a \textit{single} survey is limited by survey length. As the rate of SNe Ia is higher in more massive galaxies \citep{Mannucci2005, Sullivan2006, Graur2017b, Wiseman2021}, it is more likely that the siblings found in such single surveys are associated with high-mass galaxies ($\log(M/\mathrm{M}_{\odot})\geq10$), e.g., as found by \citet{Scolnic2020} for the siblings found in the 5-year long DES survey. This means that for cosmology, the current sample of sibling SNe Ia would predominately lie on the same side of the mass step. Additionally, more massive, passive hosts typically host faster (lower $x_1$) SNe Ia, \citep{Riess1995,Hamuy2000,Howell2001,Sullivan2010,Graur2017b,Wiseman2021}. Due to the known correlation between $x_1$ and age \citep{Howell2009,Neill2009,Johansson2013,Childress2014,Wiseman2021}, this survey bias means that the single-survey siblings are more likely to occur from older progenitors. By combining data from multiple surveys and archival data, more combinations of siblings may be found, where each sibling may have been found by a different survey. As the elapsed time between siblings will be longer in this case, siblings can be found from galaxies with lower SN Ia rates, and hence lower masses. By having more variety in duration between siblings, there may be more diversity in environmental properties.     

In this work, I search the Open Supernova Catalog \citep{OSC2017} for SNe Ia that share a host galaxy, and cross-match using the transient host galaxy database produced by \citet[][hereafter \citetalias{Qin2022}]{Qin2022}. By combining this archival sample with additional sibling SNe Ia published in the literature, I present the largest sample of sibling SNe Ia to date, which is over 10 times larger than prior sibling samples. This sample is non-exhaustive, and some siblings may have been missed if not released publicly. The sample of sibling SNe Ia is constantly increasing, so I encourage the community to use this sample, and continue to add to it as more siblings are observed.

The layout of this analysis is as follows. In \sref{sample} I discuss the creation of the sample and briefly consider some curiosities within it. In \sref{cosmology}, I identify a \lq{cosmology}\rq\ sample based on the suitability of the siblings for cosmological analyses and perform a cosmological analysis of the light-curve properties, before summarising my conclusions in \sref{conc}.

\section{Archival Sibling Identification}\label{sample}

I begin by discussing the creation of the sample. 

\subsection{The Open Supernova Catalog} \label{OSC}

First, I use the Open Supernova Catalog \citep[OSC;][]{OSC2017} to identify a sibling SNe Ia sample. The OSC was created to be a community-driven project to collect SN metadata spanning the electromagnetic spectrum, and a range of epochs of observations. This was intended to be a complete collection of public data of all known SNe, updated daily with new observations, open access and available to all in searchable catalogs on the internet. Unfortunately, in April 2022, the project ceased updating after an irrecoverable database corruption,\footnote{Status update provided on the OSC homepage: \url{https://sne.space/}} and there are no plans to resurrect the project. Despite this, the data that the project collated is freely available on the project github page and can be accessed through an API,\footnote{\url{https://github.com/astrocatalogs/OACAPI}} so provides a near-complete record of SNe up to April 2022. 

By querying the OSC API, I obtained a large sample of transients that had been classified as SNe Ia. To be a Ia, I simply require that the object has the OSC tag \texttt{claimedtype: Ia}, but this requirement is not exclusive, meaning that the objects may also have sub-classifications, for example \texttt{claimedtype: Ia Pec} or \texttt{claimedtype: Ia-91bg}. I grouped these objects by host galaxy name to find cases where a galaxy hosted more than one SN Ia (siblings). As some surveys use different naming conventions for their galaxies, I matched on all galaxy aliases to ensure that no siblings were missed. Additionally, some SN surveys use different naming conventions for their SNe; for example, a SN from the All Sky Automated Survey for SuperNovae \citep[ASAS-SN,][]{Kochanek2017} will have a different supernova designation; so they may appear on the OSC twice with different names. To combat this, I cross-checked all the aliases of the identified \lq{siblings}\rq\ on the Transient Name Server (TNS) and removed those that were duplicates.\footnote{\url{https://www.wis-tns.org/}} 

In checking the individual SNe on the TNS, I also verified the positions (ra and dec) and host galaxy association of each object, ensuring that they matched other public data, so that the sibling classification was robust. 
As an example, SN2006hj and SN2006mz are both reported on the OSC as being associated with a range of different galaxies, but have SDSS J211021.14+005557.5 in common, suggesting they may be siblings. Their ra values are very similar, but have dec values over a degree different, meaning it is very unlikely they are from the same host and this SDSS designation may have been incorrectly assigned. I therefore remove this pair from the sample.

A few objects are identified as being associated with galaxy clusters on the OSC, where the individual host galaxy within the cluster has not been determined or is disputed. As I require definitive host galaxy identification for sibling analysis, these may not be true siblings and so such objects are not included in the sample. 

Although some are considered as the likely result of a SN Ia explosion, I do not include in this sample objects that were listed on the OSC as SN remnants (SNRs). Whilst I find a number of galaxies that contain multiple SNRs (such as the Milky Way and the LMC), it is difficult to confirm without-a-doubt which are the result of a SN Ia explosion \citep[although see][for some examples of the classification of SNRs]{vandenBergh1988,Krause2008a,Krause2008b}. Therefore I do not consider them SN Ia siblings for the purposes of this sample.

Using the OSC alone, the resultant sample comprises 113 galaxies, which contain 236 confirmed SN Ia siblings.

\subsection{\citet{Qin2022} Extragalactic Transient Host Galaxy Database} \label{Qin}

Close to when the OSC ceased updating, \citetalias{Qin2022} published a database of extragalactic transients and their associated host galaxy properties.\footnote{Available at \url{https://doi.org/10.5281/zenodo.5568962}} This database cross-identifies transients with known host galaxies, and updates those for whom the host was uncertain, or unknown, with newly identified host candidates. I used this database to cross-check and to supplement my findings from the OSC. In a similar way to the OSC search, I identified transients identified as SNe Ia in this database and grouped these by host galaxy name. For all bar one sibling set, the associated host galaxy did not change, but I found a few instances where OSC had not identified a set of siblings where this database did. After a visual inspection, I agree that the hosts identified in \citetalias{Qin2022} are the likely hosts and add them to the sample. There were a few additional cases where the OSC identified a set of siblings where one SN was detected after the creation of the \citetalias{Qin2022} database, so does not appear in that sample as a sibling. 

Ignoring the cases where one SN in a sibling set was found after the creation of the \citetalias{Qin2022} database, I find three cases where siblings from the OSC are not siblings in the \citetalias{Qin2022} database. These are as follows:

\begin{enumerate}
\item If using the OSC data alone, SN2000dk and SN2015ar are labelled as originating from the same galaxy. However, after cross-checking on the TNS and in the \citetalias{Qin2022} database, this is likely not the case. In both other cases, SN2000dk is associated with NGC 382 and SN2015ar with NGC 383. NGC 382 and NGC 383 are interacting galaxies so it can be unclear which galaxy individual SNe are associated with if they are far from any one galaxy. In this case, as it is unclear if they are truly siblings or not, these objects are removed from the sample.

\item In the OSC sample, SN2010af is located in PGC 36268 with SN2004fy. However, in the \citetalias{Qin2022} database, it is labelled as also being associated with NGC 3172 along with SN2017gla. This is the only such duplicate in the sample. These three SNe are all relatively close together, but as there is considerable uncertainty over the host association it is unclear if they are siblings, so they are removed from the sample.

\item On the OSC and TNS, SN2007fq and SN2000dd are both associated with the galaxy MCG -04-48-19, where SN2007fq is close to the galaxy centre and SN2000dd is far from the galaxy light. However, in the \citetalias{Qin2022} database, which benefits from combining recent galaxy survey data, these are associated to two different, but close, galaxies. SN2007fq is reported as in ESO 528-G019 (MCG -04-48-19) and SN2000dd is located directly on a faint extended source, given the designation WISEA J203500.06-230556.9. This host may have been too faint in the original imaging of SN2000dd, hence the previous host association. As there is now doubt over the host association, this pair is removed from the sample.
\end{enumerate}

The sample from \citetalias{Qin2022} adds an additional sibling, SN2020uoo, to the existing SN2017ets-SN2020kyx pair in CGCG 137-064. In all other cases, this sample confirms existing SN Ia sibling sets from the OSC sample, or adds independent sets of SN Ia siblings to the sample.

Combining the OSC sample from \sref{OSC} with the 296 SNe Ia in 143 galaxies found in the \citetalias{Qin2022} sample gives a total of 148 galaxies containing 307 SN Ia siblings.

\subsection{Literature Siblings}

Not all surveys reported their transients to the OSC, or released them as part of a public data release that were queried by the OSC and \citetalias{Qin2022}, and instead have individual sibling sample papers. In the literature, some samples of SN Ia siblings have been collated for the purposes of cosmology or as an additional product of a large cosmological survey. Therefore, they will not have undergone the same selection cuts as outlined in this analysis. However, to boost the sample size, I add any literature SN Ia siblings that are not already present in the sample. The siblings referred to in the large \citet{Brown2014}, \citet{Burns2020} and Pantheon+ \citep{Scolnic2022} samples are found in the OSC and in the \citetalias{Qin2022} database, so the objects listed here are only those which do not:

From DES, \citet{Scolnic2020}, there are eight sets of siblings, consisting of 16 SNe Ia: 
\begin{itemize}
    \item DES13S2dlj and DES14S2pkz
    \item DES14C3zym and DES15C3edd 
    \item DES14C2iku and DES17C2jjb 
    \item DES15S2okk and DES17S2alm 
    \item DES15C2mky and DES16C2cqh 
    \item DES13E1wu and DES14E1uti 
    \item DES16C3nd$_0$ and DES16C3nd$_1$ 
    \item DES15X2mlr and DES15X2nku
\end{itemize}

From ZTF, \citet{Graham2022}, some objects are already present in the sample due to having been already added to the OSC, the \citetalias{Qin2022} database, or both. Presented here are the two sets of siblings, consisting of four SNe Ia that were not already in the sample:
\begin{itemize}
    \item SN2019sik	(ZTF19accobqx) and SN2019uej (ZTF19acnwelq)
    \item SN2019bbd	(ZTF19aaksrgj) and SN2020hzk (ZTF20aavpwxl)
\end{itemize}

\subsection{Combined Sample}\label{fullsample}

Combining the SN Ia siblings found in the OSC with those in the \citetalias{Qin2022} database, and adding those literature examples that do not appear in either, I have a final sample of \textbf{158} galaxies, containing \textbf{327} confirmed SN Ia siblings - well over 10 times larger than prior sibling samples. I present this full sample in the appendix at the end of this paper in \tref{table:allsibs} of \aref{table full sample}. 

\subsubsection{Curiosities}

I have identified a number of particularly unusual or notable SN Ia siblings within the sample. Here I give a short overview of each. 

\paragraph*{Siblings originally labelled as one object}

Typically, sibling SNe are well-separated within their host galaxies. However, there exist some recorded cases where surveys identified siblings as a single object. 

Within DES, objects found within a 1\arcsec\ radius are automatically assigned to one candidate object. However, the high redshift coverage of DES means that entire galaxies at a higher redshift may be only around 1\arcsec\ in diameter. The light-curve of one event (DES16C3nd) has been manually identified as consisting of two clear SN signatures which were separated in time by 200 days \citep{Scolnic2020}. The light curves of these objects were classified separately and were both found likely to be SNe Ia. These objects were thus given the designations DES16C3nd$_0$ and DES16C3nd$_1$. As both are SNe Ia, they are present in the sample. 

\comment{In a similar situation to DES16C3nd within the DES survey, ZTF identified a pair of sibling SNe Ia which exploded within 200 days of each other \citep{Biswas2022}. These were found at nearly identical positions, with a separation of 0.6\arcsec corresponding to a projected distance of 0.6kpc. The second transient, SN2020aewj, was classified spectroscopically as a normal SN Ia, whilst the first transient, SN2019lcj, had to be classified photometrically. As the best-fit template reported in \citet{Biswas2022} was to a normal SN Ia, this set of siblings is present in the sample.}

Alongside these, \citet{Soraisam2021} found a similar such object in the real-time alert stream of ZTF,  ZTF20aamibse (AT2020caa). This object displayed a characteristic SN light curve in 2020, with a re-brightening in 2021. The spectrum of the measured peak in 2021 has been classified as a SN Ia, however there are no spectra for the 2020 event. The authors consider that this event is likely evidence of SN siblings, but cannot conclusively determine if both of the events are SN Ia in nature. For this reason, AT2020caa is not included in the sample. 

There may be other such objects lurking in many SN surveys, and I encourage the community to carefully investigate their samples, as these peculiar objects may be unrecognised SN siblings. 

\paragraph*{Four sibling SNe Ia}

The vast majority of the sample consists of galaxies hosting a pair of siblings. Whilst there are a few galaxies that contain three SNe Ia, to date only one galaxy, NGC 1316 (Fornax A), has been found to contain four SNe Ia. These are: SN1980N, SN1981D, SN2006dd and SN2006mr, and have been extensively studied by \citet{Stritzinger2010}.

\begin{figure}
\begin{center}
\includegraphics[width=\linewidth]{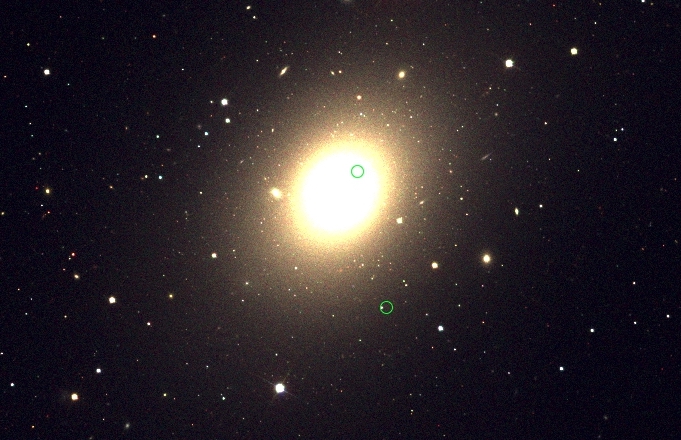}
\caption{SDSS \textit{gri} composite colour image of NGC 4636, the host galaxy in the sample hosting the siblings that are most differing in explosion date, with SN locations indicated by green circles. The SN closest to the galactic core is SN1939A, whilst SN2020ue is towards the edge of the galaxy. Imaging data obtained from SDSS Data Release 17, \citet{SDSS_DR17}.}
\label{fig:NGC4636}
\end{center}
\end{figure}

\paragraph*{Longest time between siblings}
Considering the difference in time between sets of sibling pairs, and all pairwise combinations of those consisting of three or four SNe Ia, the average time between siblings in the data is 11 years, considerably shorter than that predicted for a Milky Way-like galaxy (approximately 1 every 200 years). The longest duration between sibling SNe Ia in the sample is the 81 years between between siblings in NGC 4636: SN1939A \citep[discovered by Fritz Zwicky, one of the early pioneers of SN research,][for an overview see \citealt{Graur2022}]{Zwicky1939} and SN2020ue \citep[discovered by prolific amateur SN hunter Koichi Itagaki,][]{Itagaki2020}. I present a Sloan Digital Sky Survey (SDSS) \textit{gri} composite image of the locations of these in \fref{fig:NGC4636} to give an example sky image displaying the locations of siblings with respect to their host galaxy. As can be seen, SN1939A is located close to the nucleus of the bright NGC 4636, whilst SN2020ue is found towards the outer reach of the galaxy.

\section{Cosmology with SN Ia Siblings}\label{cosmology}

As an example use-case for this large sibling sample, I perform a brief analysis of their light-curve properties and discuss how they could be used in cosmology. 

\subsection{Photometry}

First, I obtain photometry for each object in the sample. I require that all the data be publicly available. Most objects have photometry recorded on the OSC, which can be accessed by querying the OSC API. I use this photometry as a starting point, and supplement with high-quality public ZTF photometry on the ALeRCE ZTF Explorer\footnote{\url{https://alerce.online/}}. 

Some SNe do not have publicly available photometry, whilst others do not have enough photometry to constrain a suitable model fit (in some cases, only one epoch of \lq{detection}\rq\ photometry is available). These objects are rejected from the cosmology sample before light-curve fitting. 

For the DES siblings, the photometry is not publicly available. However, these have been previously fit with SALT2 in \citet{Scolnic2020}, so I use the light-curve parameters recorded there. Additionally, the Pantheon+ siblings have already been fit in the same way by \citet{Scolnic2022}, so I can use those light-curve parameters as-well and take the averages of the multiple measurements for each Pantheon+ SN. 

\subsection{Light curve fitting} \label{lc_fitting}

In order to perform cosmology with the SN Ia siblings, the photometry needs to be of high quality so that the light curves can be accurately fit by models. I fit the light curves of the sample of sibling SNe Ia using the SALT2 fitter \citep{Guy2010} implemented in \texttt{SNCosmo} \citep{sncosmo}, with the redshifts defined as the host galaxy redshift for the shared host galaxies, \comment{consistent with the methods of \citet{Scolnic2020,Scolnic2022}}. These fits return the overall amplitude ($x_0$), the light-curve stretch ($x_1$) and the light-curve colour ($c$). As each sibling SN has been observed by different instruments, in different filters and at different times, I give SALT2 all available public data for each object, \comment{and it evaluates the appropriate wavelength range for fitting based on the redshift given for each SN, removing data outside of this range. To account for the fact that some SN have very late-time observations listed on the OSC, I apply a loose epoch cut of $<100$ days post maximum brightness (as reported on the OSC) to constrain the fit based on the requirements of SALT2. I also apply fixed Milky Way dust extinction at the SN location using the \citet{SFD1998} dust map.} 

\comment{As the fits are of varying quality for each SN given the range of photometric data on the OSC, I apply a loose quality cut where fits are rejected based on their $\chi^2$ and degrees of freedom from the SALT2 fit. To retain as many siblings as possible at this stage, I simply remove those where $\chi^2 = \textrm{inf}$ and $\textrm{dof} = 0$, indicating that the fitting has failed.}

Ignoring those siblings that appear in the DES or Pantheon+ samples, I have 103 SNe Ia which pass these loose SALT2 requirements. I can then combine with the DES and Pantheon+ samples, and from the 327 confirmed SN Ia siblings, 131 are successfully fit with SALT2, or have published SALT2 properties from DES \citep{Scolnic2020} or Pantheon+ \citep{Scolnic2022}. To these objects, I apply standard Joint Light-Curve Analysis (JLA)-like light-curve quality cuts \citep{Betoule2014}, i.e.:
\begin{itemize}
    \item $|c| < 0.3$
    \item $|x_1| < 3$
    \item $\sigma_{x_1} < 1$
\end{itemize}

This results in a total of \comment{91} SN Ia siblings\comment{, with selection cuts summarised in \tref{table:cuts}}. In some cases, one SN will pass the quality requirements, whilst its sibling does not. In order to compare properties between siblings, I require that at least two SNe from a host galaxy pass the quality cuts, resulting in \comment{21} sibling host galaxies, containing \comment{44} SNe Ia (some host galaxies contain more than two SNe Ia which pass cuts). \comment{I visually inspect all SNe that pass these cuts, and confirm that all have successful fits.} By considering all pairwise combinations, I have \comment{\textbf{25 pairs}} of siblings which pass the strict cosmology cuts. These are presented in \tref{table:cosmoparams}.

\begin{table}
\caption{\comment{Summary of selection requirements}}
\begin{threeparttable}
\begin{tabular}{cc}
\hline
\comment{Selection requirement} & \comment{Number of Supernovae}\\
\hline
\comment{Full sample} & \comment{327} \\
\comment{SALT2} & \comment{131} \\
\comment{$|c| < 0.3$} & \comment{101} \\
\comment{$|x_1| < 3$} & \comment{95} \\
\comment{$\sigma_{x_1} < 1$} & \comment{91} \\
\hline
\comment{$\ge2$ SNe per host passing requirements} & \comment{44} \\
\hline
\end{tabular}
\end{threeparttable}
\label{table:cuts}
\end{table}

\begin{table*}
\caption{Cosmology sample of SN Ia Siblings.}
\begin{threeparttable}
\begin{tabular}{lcccccc}
\hline
\hline
Host name &     Host $z$ &              Sibling SN &     $x_1 \pm \sigma_{x_1}$ &      $c \pm  \sigma_c$ &     $m_B \pm  \sigma_{m_B}$ & SALT2 fit\\
\hline
\hline
NGC 1448 & 0.004 &              SN2001el & -0.130 $\pm$   0.030 &  0.070 $\pm$  0.030 & 12.420 $\pm$   0.030 & Pantheon+ \\
 &  &             SN2021pit & -0.040 $\pm$   0.120 &  0.070 $\pm$  0.040 & 12.030 $\pm$   0.040 & Pantheon+ \\
\hline
NGC 5643 & 0.005 &              SN2013aa &  0.555 $\pm$   0.180 & -0.130 $\pm$  0.064 & 10.825 $\pm$   0.149 & Pantheon+ \\
 &  &             SN2017cbv &  0.710 $\pm$   0.081 & -0.120 $\pm$  0.050 & 10.770 $\pm$   0.135 & Pantheon+ \\
\hline
NGC 1316 & 0.006 &               SN1980N & -1.140 $\pm$   0.120 & -0.010 $\pm$  0.030 & 12.090 $\pm$   0.030 & Pantheon+ \\
 &  &               SN1981D & -1.150 $\pm$   0.360 &  0.160 $\pm$  0.060 & 12.220 $\pm$   0.060 & Pantheon+ \\
 &  &              SN2006dd & -0.290 $\pm$   0.040 &  0.020 $\pm$  0.030 & 11.990 $\pm$   0.030 & Pantheon+ \\
\hline
NGC 1404 & 0.006 &              SN2007on & -2.220 $\pm$   0.040 &  0.000 $\pm$  0.020 & 12.700 $\pm$   0.020 & Pantheon+ \\
 &  &              SN2011iv & -1.900 $\pm$   0.040 & -0.060 $\pm$  0.020 & 12.110 $\pm$   0.020 & Pantheon+ \\
\hline
NGC 5018 & 0.009 & SN2002dj & -0.211 $\pm$	0.129	& -0.010 $\pm$	0.005	& 13.933	$\pm$ 0.014 & This work \\
& & SN2021fxy & 0.135 $\pm$	0.163 &	-0.286	$\pm$ 0.007	& 13.422	$\pm$ 0.014 & This work \\
\hline
NGC 3147 & 0.009 &              SN1997bq & -0.610 $\pm$   0.090 &  0.080 $\pm$  0.030 & 14.100 $\pm$   0.040 & Pantheon+ \\
 &  &              SN2008fv &  0.740 $\pm$   0.060 &  0.110 $\pm$  0.030 & 14.220 $\pm$   0.030 & Pantheon+ \\
 &  &             SN2021hpr &  0.250 $\pm$   0.070 &  0.040 $\pm$  0.030 & 13.980 $\pm$   0.030 & Pantheon+ \\
\hline
NGC 5468 & 0.010 &              SN1999cp &  0.155 $\pm$   0.089 & -0.080 $\pm$  0.050 & 13.625 $\pm$   0.042 & Pantheon+ \\
 &  &              SN2002cr & -0.485 $\pm$   0.067 & -0.055 $\pm$  0.036 & 13.890 $\pm$   0.036 & Pantheon+ \\
\hline
NGC 6956 & 0.014 &              SN2013fa & -0.560 $\pm$   0.090 &  0.200 $\pm$  0.030 & 15.320 $\pm$   0.060 & Pantheon+ \\
 &  & PSN J20435314+1230304 & -2.550 $\pm$   0.150 &  0.090 $\pm$  0.040 & 15.680 $\pm$   0.070 & Pantheon+ \\
\hline
NGC 7311	& 0.015 &	SN2005kc	& -0.706	$\pm$ 0.040	& 0.206	$\pm$ 0.002	& 15.549	$\pm$ 0.002 & This work \\
& &	SN2020adcb	& -0.800	$\pm$ 0.167	& 0.056	$\pm$ 0.010	& 15.657	$\pm$ 0.013 & This work\\
\hline
NGC 4493 & 0.023 &               SN1994M & -1.430 $\pm$   0.090 &  0.040 $\pm$  0.030 & 15.980 $\pm$   0.040 & Pantheon+ \\
 &  &              SN2004br &  1.000 $\pm$   0.080 & -0.110 $\pm$  0.030 & 15.120 $\pm$   0.030 & Pantheon+ \\
\hline
NGC 6240 & 0.024 &              SN2010gp & -0.510 $\pm$   0.250 &  0.170 $\pm$  0.050 & 15.890 $\pm$   0.060 & Pantheon+ \\
 &  &              PS1-14xw &  1.230 $\pm$   0.710 & -0.040 $\pm$  0.060 & 15.570 $\pm$   0.060 & Pantheon+ \\
\hline
UGC 7228 & 0.026 &              SN2007sw &  0.080 $\pm$   0.130 &  0.090 $\pm$  0.030 & 15.980 $\pm$   0.040 & Pantheon+ \\
 &  &              SN2012bh & -0.410 $\pm$   0.260 & -0.040 $\pm$  0.050 & 15.920 $\pm$   0.090 & Pantheon+ \\
\hline
UGC 3218	& 0.026	& SN2006le	& 0.902	$\pm$ 0.289	& -0.095	$\pm$ 0.024	& 14.998	$\pm$ 0.022 & This work \\
	& 	& SN2011M	& -0.436	$\pm$ 0.107	& -0.019	$\pm$ 0.007	& 15.248	$\pm$ 0.009 & This work \\
\hline
NGC 1575	& 0.031	& SN2020sjo	& -0.681	$\pm$ 0.099	& -0.197	$\pm$ 0.013	& 16.403	$\pm$ 0.017 & This work \\
	& 	& SN2020zhh	& -0.227	$\pm$ 0.069	& -0.167	$\pm$ 0.008	& 16.287	$\pm$ 0.009& This work \\
\hline
WISEA J135842.24+430727.7	& 0.072	& SN2020jdq	& -1.125	$\pm$ 0.355	& -0.289	$\pm$ 0.045	& 18.548	$\pm$ 0.039 & This work \\
	& 	& SN2021meh	& -0.759	$\pm$ 0.565	& -0.190	$\pm$ 0.046	& 18.552	$\pm$ 0.054 & This work \\
\hline
WISEA J123806.26+080245.2	& 0.086	& SN2019bbd	& 2.106	$\pm$ 0.760	& -0.242	$\pm$ 0.045	& 18.652	$\pm$ 0.042 & This work \\
	& 	& SN2020hzk	& 1.551	$\pm$ 0.936	& 0.092	$\pm$ 0.061	& 19.050	$\pm$ 0.078 & This work \\
\hline
 & 0.228 &            DES13S2dlj &  0.300 $\pm$   0.230 &  0.186 $\pm$  0.030 & 21.610 $\pm$   0.040 & DES\\
 &  &            DES14S2pkz & -0.407 $\pm$   0.140 &  0.080 $\pm$  0.020 & 21.110 $\pm$   0.030 & DES\\
\hline
 & 0.349 &            DES14C3zym & -1.560 $\pm$   0.150 & -0.070 $\pm$  0.030 & 22.050 $\pm$   0.030 & DES\\
 &  &            DES15C3edd & -1.370 $\pm$   0.160 & -0.050 $\pm$  0.030 & 22.180 $\pm$   0.030 & DES\\
\hline
 & 0.506 &            DES15S2okk &  0.089 $\pm$   0.650 & -0.040 $\pm$  0.030 & 22.550 $\pm$   0.040 & DES\\
 &  &            DES17S2alm &  1.220 $\pm$   0.620 & -0.050 $\pm$  0.040 & 22.890 $\pm$   0.060 & DES\\
\hline
 & 0.524 &            DES15C2mky &  0.820 $\pm$   0.340 & -0.080 $\pm$  0.030 & 22.640 $\pm$   0.040 & DES\\
 &  &            DES16C2cqh & -0.847 $\pm$   0.810 &  0.020 $\pm$  0.020 & 23.520 $\pm$   0.080 & DES\\
\hline
 & 0.648 &            DES16C3nd0 & -1.080 $\pm$   0.290 &  0.010 $\pm$  0.040 & 23.520 $\pm$   0.030 & DES\\
 &  &            DES16C3nd1 & -0.452 $\pm$   0.280 &  0.040 $\pm$  0.030 & 23.340 $\pm$   0.030 & DES\\
\hline
\hline
\end{tabular}
\begin{tablenotes}
\item Pantheon+ refers to \citet{Scolnic2022}, and DES to \citet{Scolnic2020}.
\end{tablenotes}
\end{threeparttable}
\label{table:cosmoparams}
\end{table*}

\subsection{Light-curve properties} \label{lc_props}

\begin{figure}
\begin{center}
\includegraphics[width=0.90\linewidth]{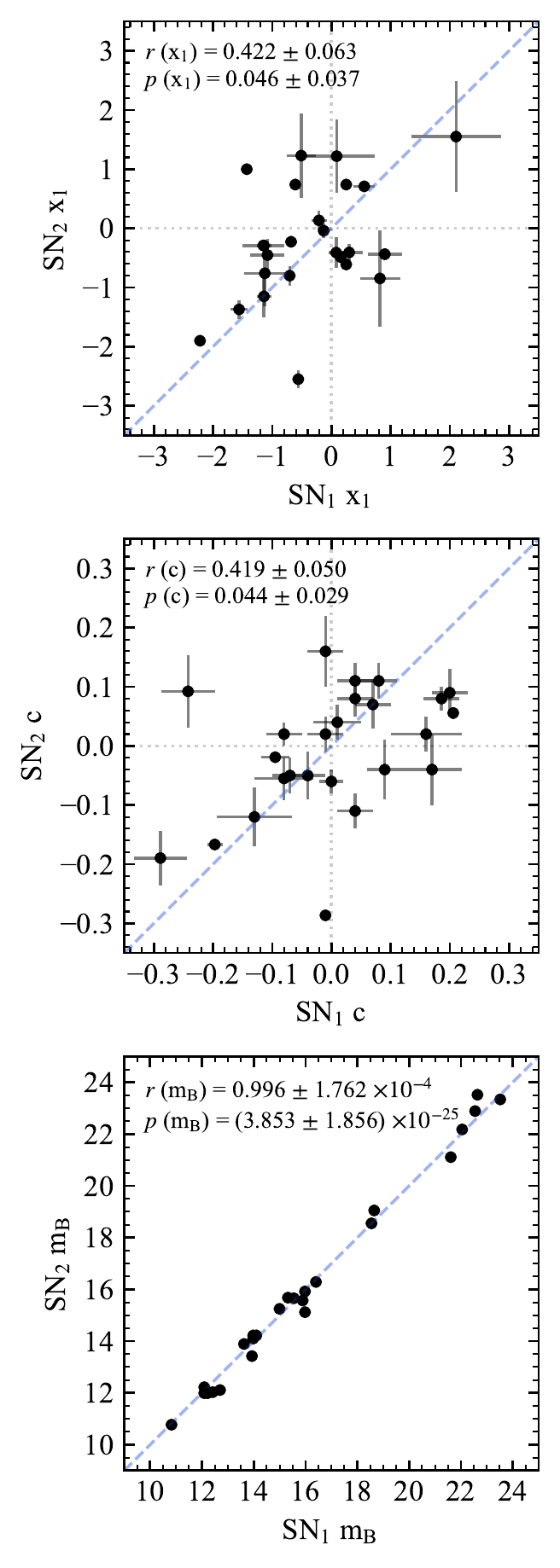}
\caption{Comparison of light-curve parameters $x_1$ (top panel), $c$ (middle panel) and m$_B$ (lower panel) for all pairs of SN siblings in the cosmology sample. The blue dashed line represents the $x=y$ line where sibling light-curve parameters are equal. Presented in the top left corner for each plot are the Pearson $r$ values for the data, the associated $p$-values, and uncertainties on both.}
\label{fig:sibs_lcprops_comp}
\end{center}
\end{figure}

I first compare the differences in light-curve properties $x_1$, $c$ and peak B-band apparent magnitude ($\textrm{m}_\textrm{B}$) for all pairs of SN siblings in the cosmology sample. These comparisons are presented in \fref{fig:sibs_lcprops_comp}, with Pearson $r$ values to interpret potential correlations in the data. As there are uncertainties in both the x and y axes, I use a Monte Carlo technique to account for the uncertainties in the correlation. By taking the x and y uncertainties around each data point as a sampling area, I obtain a random location within the defined area for each SN pair and calculate the $r$ value. I then apply this technique 10,000 times, taking the mean and standard deviation of the calculated $r$ values as the defined $r$ value and uncertainty on $r$ for the data. This is repeated for $x_1$, $c$ and $\textrm{m}_\textrm{B}$.

As shown in \fref{fig:sibs_lcprops_comp}, the only quantity that is correlated between siblings is the apparent magnitude $\textrm{m}_\textrm{B}$, with \comment{$r = 0.995 \pm 1.762 \times 10^{-4}$ indicating a very strong positive correlation, and a $p$-value $= (3.853 \pm 1.856) \times 10^{-25}$,} meaning that the null hypothesis can be rejected and the correlation is highly significant. This is to be expected as $\textrm{m}_\textrm{B}$ is primarily dependent on the distance to the SN (or the redshift), so should be correlated for SNe at the same redshift. 

\comment{For $x_1$ and $c$, it can be shown that there is considerable scatter, and the $r$-value and $p$-values for each are comparable. For $x_1$, the $r$-value is measured at $r = 0.422 \pm 0.063$. Whilst this may be considered to be a weak positive correlation, the $p$-value of $0.046 \pm 0.037$ suggests that this is not significant and the x and y data (i.e., the $x_1$ values between sibling pairs) may be unrelated, assuming a typical $p$-value threshold of 0.01. This is also similar for $c$, with $r = 0.419 \pm 0.050$ and $p = 0.044 \pm 0.029$; meaning that the $c$ values between sibling pairs may be unrelated. A larger sample of cosmological siblings is needed to further investigate this.} 

\begin{figure}
\begin{center}
\includegraphics[width=\linewidth]{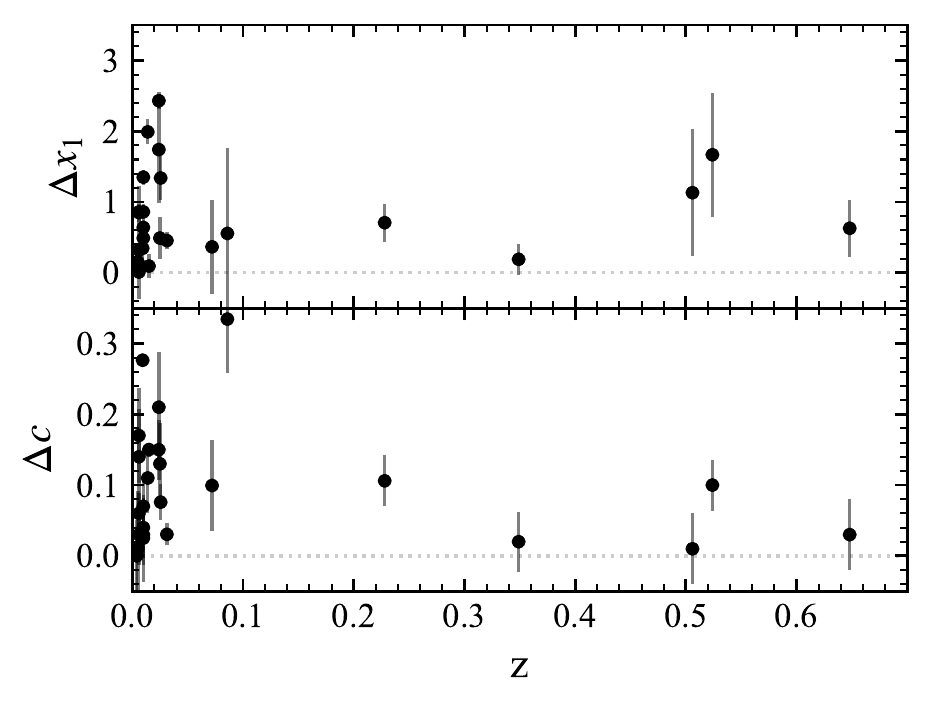}
\caption{Differences in light-curve parameters $x_1$ (top panel) and $c$ (lower panel), as a function of redshift for all pairs of SN siblings in the cosmology sibling sample.}
\label{fig:sibs_lcprops_vs_z}
\end{center}
\end{figure}

Given that I have cosmological siblings that span a large redshift range, it is prudent to investigate potential trends with redshift. Thus, in \fref{fig:sibs_lcprops_vs_z}, I present the differences in $x_1$ and $c$ between sibling pairs ($\Delta x_1$ and $\Delta c$) as a function of redshift. No clear trends are evident, but I note that the overall scatter in $\Delta x_1$ and $\Delta c$ seems to reduce towards higher redshifts. This however is likely due to small number statistics, as I only have five pairs of SNe with $z>0.1$. This may also be due to Malmquist bias, as at high redshift we are more likely to observe the brighter, high-stretch SNe.

\begin{figure*}
\begin{center}
    \begin{subfigure}[b]{0.49\textwidth}
    \includegraphics[width=\linewidth]{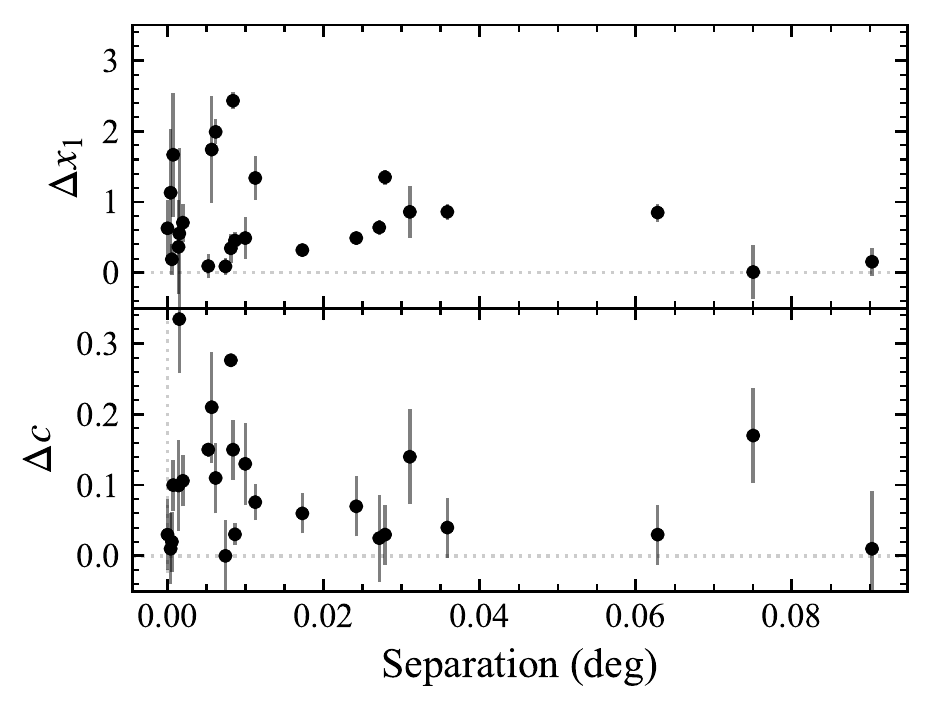}
    \caption{}
    \label{fig:vssep}
    \end{subfigure}
    \begin{subfigure}[b]{0.49\textwidth}
    \includegraphics[width=\linewidth]{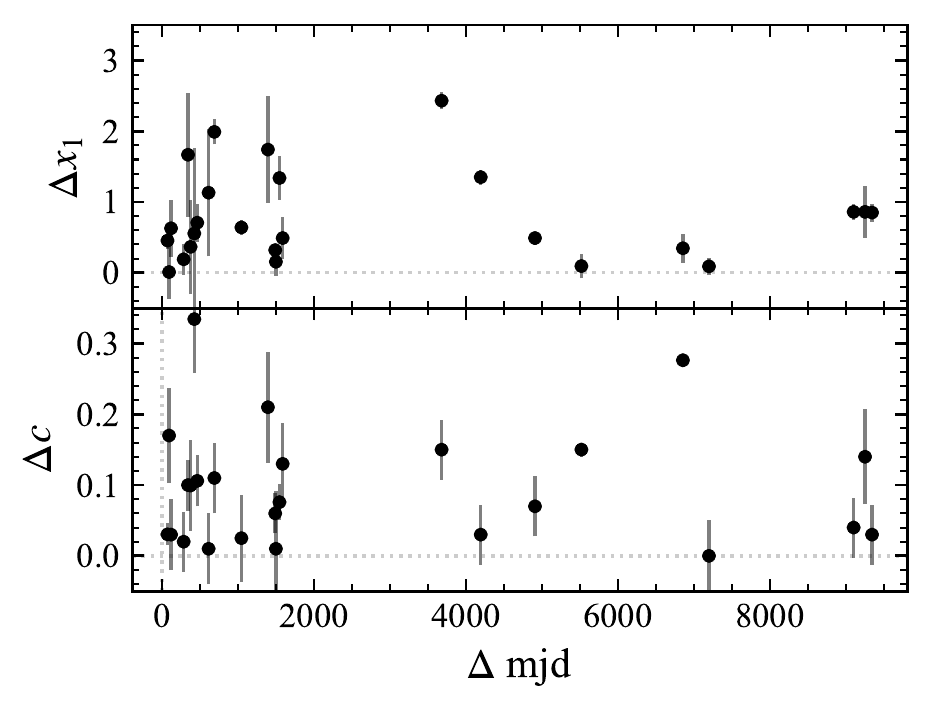}
    \caption{}
    \label{fig:vsmjd}
    \end{subfigure}

\caption{Differences in light-curve parameters $x_1$ (top panel) and $c$ (lower panel), as a function of (a) angular separation between SNe for all pairs of SN siblings in the cosmology sample, and (b) the difference in the date of peak magnitude for all pairs of SN siblings in the cosmology sample.}
\label{fig:sibs_lcprops_vs_angandmjd}
\end{center}
\end{figure*}

I also investigate potential trends with the angular separation between sibling pairs, and with the difference in the date of peak magnitude between sibling pairs, finding no evidence of any trend for either, as shown in \fref{fig:sibs_lcprops_vs_angandmjd}. 

The scatter in $\Delta x_1$, $\Delta c$ and $\Delta \textrm{m}_\textrm{B}$ provides an interesting way to compare siblings to a sample of random pairs of SNe, as in \citet{Scolnic2022}. I find the standard deviations ($\sigma$) of the $\Delta x_1$, $\Delta c$ and $\Delta \textrm{m}_\textrm{B}$ values and present them in \tref{table:simulation_comparison}. 

To compare the results obtained for SN Ia siblings to a standard sample of SNe Ia (i.e., non-siblings), I perform a simple \comment{boot-strapping analysis}. I take the sample of SNe Ia from Pantheon+ \citep{Scolnic2022}, make a cut on redshift to match the siblings in my cosmology sample, and apply the same JLA-like light-curve quality cuts from \sref{lc_fitting}. This gives 1418 SNe Ia to sample from. I select \comment{25} pairs of random SNe from this sample, and repeat the Monte Carlo analysis to obtain Pearson $r$ values and to calculate scatter in light-curve properties. I repeat this 10,000 times and obtain the means and standard deviations of the data. \comment{I study the effect of performing this boot-strapping analysis on smaller redshift sub-samples in \aref{further_bootstrap}.}

For the $r$ values comparing light-curve properties between random pairs of SNe, I find that there are consistently no correlations for the different properties. In detail, I find for \comment{$x_1$ the $r$-value is measured at $r = 0.002 \pm 0.057$ with a $p$-value of $0.503 \pm 0.137$; for $c$, $r = 0.004 \pm 0.073$ with a $p$-value of $0.497 \pm 0.167$; and for $\textrm{m}_\textrm{B}$, $r = 0.001 \pm 0.003$ with a $p$-value of $0.501 \pm 0.008$}. Whilst the lack of any correlation suggests that one could argue that sibling SNe are more correlated than any random pair of SNe, the associated $p$-values are consistent with a lack of correlation across both data and \comment{the boot-strap sample}. Therefore I can only confidently say that the $\textrm{m}_\textrm{B}$ values for SN siblings are more similar than any other random pair of SNe, which is to be expected. 

Comparing between data and \comment{boot-strapped}\ standard deviations ($\sigma$) in the differences in light-curve properties between sibling pairs can provide more answers. Presented in \tref{table:simulation_comparison} are the different $\sigma$ values for the sibling data and for any random pair of SNe. As can be seen, the \comment{boot-strapped}\ $\sigma$ values are higher than that for the data, with the most significant difference (at $6\sigma$ significance) being for the difference in scatter in $\Delta \textrm{m}_\textrm{B}$, which will be due to the siblings being co-located in the same galaxy. For $\Delta x_1$ and $\Delta c$, the differences in scatter are \comment{$<2\sigma$ different}. Therefore, I can say that $x_1$ and $c$ are no more similar between SNe Ia associated with the same host galaxy than for any random pair of SNe Ia. This finding raises important questions over the validity of global host galaxy corrections in the standardisation of SNe Ia for use in cosmology.

\begin{table}
\caption{Standard deviation $\sigma$ of the differences in light-curve properties between SN$_\textrm{1}$ and SN$_\textrm{2}$ for the data and for a \comment{sample obtained through boot-strapping}.}
\begin{threeparttable}
\makebox[1 \linewidth][c]{       %fixing table spacing
\begin{tabular}{cccc}
\hline
Property &     Data $\sigma$ &              \comment{Boot-strap}\ $\sigma$ &     Significance of difference\\
\hline
$\Delta x_1$ & $0.629 \pm 0.091$ & $0.809 \pm 0.117$ & $1.215$\\
$\Delta c$ & $0.084 \pm 0.012$ & $0.072 \pm 0.010$ & $0.733$\\
$\Delta \textrm{m}_\textrm{B}$ & $0.228 \pm 0.033$ & $2.468 \pm 0.356$ & $6.260$\\
\hline
\end{tabular}
}
\end{threeparttable}
\label{table:simulation_comparison}
\end{table}

This finding for $x_1$ disagrees with \citet{Scolnic2020, Scolnic2022}, which both find tentative evidence that $x_1$ is more correlated between siblings than random pairs of SNe. This difference may be due to different light-curve quality cuts in our samples, or their smaller sample size. Like \citet{Scolnic2020,Scolnic2022} I find no evidence that $c$ is more similar in sibling SNe Ia than in any random pair of SNe Ia. Further investigation of this difference is needed. 

\comment{Assuming a simple reference cosmology of $\alpha = 0.148$ and $\beta = 3.112$ based on the values for Pantheon+ in \citet{Scolnic2022}, the standard deviation on the difference in distance modulus between siblings in this analysis is $0.185$ mag. This is less than the $0.32$ mag value found by \citet{Scolnic2022} for the Pantheon+ siblings alone, but is still more than the $\sim 0.15$ mag dispersion on the Hubble diagram \citep[e.g.][]{Brout2022}, agreeing with the conclusions that siblings are no more similar than any random pair of SNe Ia. This is in contrast to studies such as \citet{Burns2020}, who find that the relative distances between siblings agreed within 3\%. However, these were all spectroscopically confirmed SNe Ia with well-sampled light-curves, consistent photometry and data for sets of siblings from the same telescopes. Whereas the sample compiled by my analysis comprises a mixture of those features. In particular, \citet{Burns2020} note that their largest distance discrepancies ($5-10\%$) arise when siblings are observed with different telescopes. This may be due to systematics in the photometric systems, but may also be due to differences in the SNe themselves. As postulated in \sref{intro}, to have been observed by the same survey, it is more likely that the SNe Ia will have come from a more massive galaxy, hosting faster SNe Ia, from older progenitors. Contrasting with \citet{Burns2020}, in this analysis the siblings with the most differing distance modulii ($0.725$ mag) are SN2019bbd and SN2020hzk, both observed by ZTF. Despite having similar $x_1$ values, the differences in $c$ between the siblings is high, suggesting perhaps differing levels of dust attenuation or differing progenitor properties, as also found by \citet{Elias-Rosa2008} and \citet{Biswas2022} for other sets of siblings. Future analysis is needed to understand how much of this difference across sets of siblings is an environmental effect, and how much is a systematic from combining different telescope data. Regardless of the cause, understanding these differences between SNe Ia from the same galaxy is vital as we enter the new era of precision cosmology.}

\section{Discussion and Conclusions}\label{conc}

In this analysis I have identified the largest sample of SN Ia siblings to date by searching the Open Supernova Catalog, the \citetalias{Qin2022} Extragalactic Transient Host Galaxy Database, and the literature. This results in a combined sample of \textbf{158} galaxies, containing \textbf{327} SN Ia siblings, over an order of magnitude larger than prior samples. 

By creating a sub-sample based on photometry quality and applying JLA-like light-curve quality cuts, I identified a cosmology sample of \comment{25} pairs of siblings, which I used to compare light-curve properties between sets of siblings. I found that the only property that was correlated between siblings was the peak apparent magnitude $\textrm{m}_{\textrm{B}}$, which is expected due to sets of sibling SNe being at the same redshift. There was no correlation in either $x_1$ or $c$ for SNe associated with the same host galaxy, suggesting that these properties are not consistent for SNe Ia associated with the same galaxies. 

Using a sample of non-sibling SNe Ia from the Pantheon+ sample, I compare the siblings to random pairs of SNe Ia, finding that $x_1$ and $c$ values for siblings are no more similar than for any random pair of SNe Ia. In the current cosmology formalism, environmental effects are corrected for by considering the mass of the galaxy that each SN is associated with, meaning that sibling SNe Ia are subject to the same corrections. My finding that their $x_1$ and $c$ values are no more similar than any random pair of SNe raises important questions about the validity of this method, as this global host galaxy correction does not account for sub-galactic environmental differences. The local environments of sibling SNe Ia should be investigated and compared to light-curve properties to understand if using local corrections may better account for environmental property standardisation in cosmology as suggested by \citet{Kelsey2021,Kelsey2022}. It is clear that SN siblings are a vital tool to understand the remaining dispersion in SN Ia luminosities. 

As we enter the new era of large surveys, the number of SN Ia siblings will dramatically increase. \citet{Scolnic2020}, for example, predict around 800 pairs in LSST, alongside the future matching of newly discovered SNe with archival SNe sharing the same host galaxy. This will allow for further analysis of differences between sibling pairs, and for the detailed investigation of their environments. 

\section*{Acknowledgements}

I thank the UKRI Future Leaders Fellowship for support through the grant MR/T01881X/1. I thank Or Graur and Laura Nuttall for valuable comments on the manuscript and also thank Peter Clark and Mark Magee for the useful discussions.

I thank the Open Supernova Catalog team for making the data publicly available on GitHub \citet{OSC2017}, and to \citetalias{Qin2022} for making their database publicly available on Zenodo. 

This analysis used \textsc{pandas} \citep{McKinney2010}, \textsc{numpy} \citep{Harris2020}, \textsc{scipy} \citep{Virtanen2020}, \textsc{astropy} \citep{astropy:2018}, \textsc{sncosmo} \citep{sncosmo}, and \textsc{matplotlib} \citep{Hunter2007}.

%%%%%%%%%%%%%%%%%%%%%%%%%%%%%%%%%%%%%%%%%%%%%%%%%%
\section*{Data Availability}

The new data underlying this article are available in the article and associated appendices. This work is based on data that are publicly available on the Open Supernova Catalog \citep{OSC2017}, \citetalias{Qin2022} and \citet{Scolnic2022}. 



\bibliographystyle{mnras}
\bibliography{biblio} % if your bibtex file is called example.bib

\begin{thebibliography}{}
\makeatletter
\relax
\def\mn@urlcharsother{\let\do\@makeother \do\$\do\&\do\#\do\^\do\_\do\%\do\~}
\def\mn@doi{\begingroup\mn@urlcharsother \@ifnextchar [ {\mn@doi@}
  {\mn@doi@[]}}
\def\mn@doi@[#1]#2{\def\@tempa{#1}\ifx\@tempa\@empty \href
  {http://dx.doi.org/#2} {doi:#2}\else \href {http://dx.doi.org/#2} {#1}\fi
  \endgroup}
\def\mn@eprint#1#2{\mn@eprint@#1:#2::\@nil}
\def\mn@eprint@arXiv#1{\href {http://arxiv.org/abs/#1} {{\tt arXiv:#1}}}
\def\mn@eprint@dblp#1{\href {http://dblp.uni-trier.de/rec/bibtex/#1.xml}
  {dblp:#1}}
\def\mn@eprint@#1:#2:#3:#4\@nil{\def\@tempa {#1}\def\@tempb {#2}\def\@tempc
  {#3}\ifx \@tempc \@empty \let \@tempc \@tempb \let \@tempb \@tempa \fi \ifx
  \@tempb \@empty \def\@tempb {arXiv}\fi \@ifundefined
  {mn@eprint@\@tempb}{\@tempb:\@tempc}{\expandafter \expandafter \csname
  mn@eprint@\@tempb\endcsname \expandafter{\@tempc}}}

\bibitem[\protect\citeauthoryear{{Abdurro'uf} et~al.,}{{Abdurro'uf}
  et~al.}{2022}]{SDSS_DR17}
{Abdurro'uf} et~al., 2022, \mn@doi [\apjs] {10.3847/1538-4365/ac4414}, \href
  {https://ui.adsabs.harvard.edu/abs/2022ApJS..259...35A} {259, 35}

\bibitem[\protect\citeauthoryear{{Anderson} \& {Soto}}{{Anderson} \&
  {Soto}}{2013}]{AndersonSoto2013}
{Anderson} J.~P.,  {Soto} M.,  2013, \mn@doi [\aap]
  {10.1051/0004-6361/201220600}, \href
  {https://ui.adsabs.harvard.edu/abs/2013A&A...550A..69A} {550, A69}

\bibitem[\protect\citeauthoryear{{Ashall} et~al.,}{{Ashall}
  et~al.}{2018}]{Ashall2018}
{Ashall} C.,  et~al., 2018, \mn@doi [\mnras] {10.1093/mnras/sty632}, \href
  {https://ui.adsabs.harvard.edu/abs/2018MNRAS.477..153A} {477, 153}

\bibitem[\protect\citeauthoryear{Barbary et~al.,}{Barbary
  et~al.}{2022}]{sncosmo}
Barbary K.,  et~al., 2022, SNCosmo, \mn@doi{10.5281/zenodo.7117347}, \url
  {https://doi.org/10.5281/zenodo.7117347}

\bibitem[\protect\citeauthoryear{{Betoule} et~al.,}{{Betoule}
  et~al.}{2014}]{Betoule2014}
{Betoule} M.,  et~al., 2014, \mn@doi [\aap] {10.1051/0004-6361/201423413},
  \href {https://ui.adsabs.harvard.edu/abs/2014A&A...568A..22B} {568, A22}

\bibitem[\protect\citeauthoryear{{Biswas} et~al.,}{{Biswas}
  et~al.}{2022}]{Biswas2022}
{Biswas} R.,  et~al., 2022, \mn@doi [\mnras] {10.1093/mnras/stab2943}, \href
  {https://ui.adsabs.harvard.edu/abs/2022MNRAS.509.5340B} {509, 5340}

\bibitem[\protect\citeauthoryear{{Brout} et~al.,}{{Brout}
  et~al.}{2022}]{Brout2022}
{Brout} D.,  et~al., 2022, \mn@doi [\apj] {10.3847/1538-4357/ac8e04}, \href
  {https://ui.adsabs.harvard.edu/abs/2022ApJ...938..110B} {938, 110}

\bibitem[\protect\citeauthoryear{{Brown}}{{Brown}}{2014}]{Brown2014}
{Brown} P.,  2014, in Proceedings of Swift: 10 Years of Discovery (SWIFT 10).
  p.~125 (\mn@eprint {arXiv} {1505.01368})

\bibitem[\protect\citeauthoryear{{Burns} et~al.,}{{Burns}
  et~al.}{2020}]{Burns2020}
{Burns} C.~R.,  et~al., 2020, \mn@doi [\apj] {10.3847/1538-4357/ab8e3e}, \href
  {https://ui.adsabs.harvard.edu/abs/2020ApJ...895..118B} {895, 118}

\bibitem[\protect\citeauthoryear{{Childress} et~al.,}{{Childress}
  et~al.}{2013}]{Childress2013}
{Childress} M.,  et~al., 2013, \mn@doi [\apj] {10.1088/0004-637X/770/2/108},
  \href {https://ui.adsabs.harvard.edu/abs/2013ApJ...770..108C} {770, 108}

\bibitem[\protect\citeauthoryear{{Childress}, {Wolf}  \& {Zahid}}{{Childress}
  et~al.}{2014}]{Childress2014}
{Childress} M.~J.,  {Wolf} C.,   {Zahid} H.~J.,  2014, \mn@doi [\mnras]
  {10.1093/mnras/stu1892}, \href
  {https://ui.adsabs.harvard.edu/abs/2014MNRAS.445.1898C} {445, 1898}

\bibitem[\protect\citeauthoryear{{D'Andrea} et~al.,}{{D'Andrea}
  et~al.}{2011}]{DAndrea2011}
{D'Andrea} C.~B.,  et~al., 2011, \mn@doi [\apj] {10.1088/0004-637X/743/2/172},
  \href {https://ui.adsabs.harvard.edu/abs/2011ApJ...743..172D} {743, 172}

\bibitem[\protect\citeauthoryear{{DerKacy} et~al.,}{{DerKacy}
  et~al.}{2023}]{DerKacy2022}
{DerKacy} J.~M.,  et~al., 2023, \mn@doi [\mnras] {10.1093/mnras/stad1171},
  \href {https://ui.adsabs.harvard.edu/abs/2023MNRAS.522.3481D} {522, 3481}

\bibitem[\protect\citeauthoryear{{Elias-Rosa} et~al.,}{{Elias-Rosa}
  et~al.}{2008}]{Elias-Rosa2008}
{Elias-Rosa} N.,  et~al., 2008, \mn@doi [\mnras]
  {10.1111/j.1365-2966.2007.12638.x}, \href
  {https://ui.adsabs.harvard.edu/abs/2008MNRAS.384..107E} {384, 107}

\bibitem[\protect\citeauthoryear{{Gall} et~al.,}{{Gall}
  et~al.}{2018}]{Gall2018}
{Gall} C.,  et~al., 2018, \mn@doi [\aap] {10.1051/0004-6361/201730886}, \href
  {https://ui.adsabs.harvard.edu/abs/2018A&A...611A..58G} {611, A58}

\bibitem[\protect\citeauthoryear{{Gallego-Cano}, {Izzo}, {Dominguez-Tagle},
  {Prada}, {P{\'e}rez}, {Khetan}  \& {Jang}}{{Gallego-Cano}
  et~al.}{2022}]{Gallego-Cano2022}
{Gallego-Cano} E.,  {Izzo} L.,  {Dominguez-Tagle} C.,  {Prada} F.,  {P{\'e}rez}
  E.,  {Khetan} N.,   {Jang} I.~S.,  2022, \mn@doi [\aap]
  {10.1051/0004-6361/202243988}, \href
  {https://ui.adsabs.harvard.edu/abs/2022A&A...666A..13G} {666, A13}

\bibitem[\protect\citeauthoryear{{Graham} et~al.,}{{Graham}
  et~al.}{2022}]{Graham2022}
{Graham} M.~L.,  et~al., 2022, \mn@doi [\mnras] {10.1093/mnras/stab3802}, \href
  {https://ui.adsabs.harvard.edu/abs/2022MNRAS.511..241G} {511, 241}

\bibitem[\protect\citeauthoryear{{Graur}}{{Graur}}{2022}]{Graur2022}
{Graur} O.,  2022, {Supernova}

\bibitem[\protect\citeauthoryear{{Graur}, {Bianco}, {Huang}, {Modjaz},
  {Shivvers}, {Filippenko}, {Li}  \& {Eldridge}}{{Graur}
  et~al.}{2017a}]{Graur2017a}
{Graur} O.,  {Bianco} F.~B.,  {Huang} S.,  {Modjaz} M.,  {Shivvers} I.,
  {Filippenko} A.~V.,  {Li} W.,   {Eldridge} J.~J.,  2017a, \mn@doi [\apj]
  {10.3847/1538-4357/aa5eb8}, \href
  {https://ui.adsabs.harvard.edu/abs/2017ApJ...837..120G} {837, 120}

\bibitem[\protect\citeauthoryear{{Graur}, {Bianco}, {Modjaz}, {Shivvers},
  {Filippenko}, {Li}  \& {Smith}}{{Graur} et~al.}{2017b}]{Graur2017b}
{Graur} O.,  {Bianco} F.~B.,  {Modjaz} M.,  {Shivvers} I.,  {Filippenko} A.~V.,
   {Li} W.,   {Smith} N.,  2017b, \mn@doi [\apj] {10.3847/1538-4357/aa5eb7},
  \href {https://ui.adsabs.harvard.edu/abs/2017ApJ...837..121G} {837, 121}

\bibitem[\protect\citeauthoryear{{Guillochon}, {Parrent}, {Kelley}  \&
  {Margutti}}{{Guillochon} et~al.}{2017}]{OSC2017}
{Guillochon} J.,  {Parrent} J.,  {Kelley} L.~Z.,   {Margutti} R.,  2017,
  \mn@doi [\apj] {10.3847/1538-4357/835/1/64}, \href
  {https://ui.adsabs.harvard.edu/abs/2017ApJ...835...64G} {835, 64}

\bibitem[\protect\citeauthoryear{{Gupta} et~al.,}{{Gupta}
  et~al.}{2011}]{Gupta2011}
{Gupta} R.~R.,  et~al., 2011, \mn@doi [\apj] {10.1088/0004-637X/740/2/92},
  \href {https://ui.adsabs.harvard.edu/abs/2011ApJ...740...92G} {740, 92}

\bibitem[\protect\citeauthoryear{{Guy} et~al.,}{{Guy} et~al.}{2010}]{Guy2010}
{Guy} J.,  et~al., 2010, \mn@doi [\aap] {10.1051/0004-6361/201014468}, \href
  {https://ui.adsabs.harvard.edu/abs/2010A&A...523A...7G} {523, A7}

\bibitem[\protect\citeauthoryear{{Hamuy}, {Phillips}, {Maza}, {Wischnjewsky},
  {Uomoto}, {Landolt}  \& {Khatwani}}{{Hamuy} et~al.}{1991}]{Hamuy1991}
{Hamuy} M.,  {Phillips} M.~M.,  {Maza} J.,  {Wischnjewsky} M.,  {Uomoto} A.,
  {Landolt} A.~U.,   {Khatwani} R.,  1991, \mn@doi [\aj] {10.1086/115867},
  \href {https://ui.adsabs.harvard.edu/abs/1991AJ....102..208H} {102, 208}

\bibitem[\protect\citeauthoryear{{Hamuy}, {Trager}, {Pinto}, {Phillips},
  {Schommer}, {Ivanov}  \& {Suntzeff}}{{Hamuy} et~al.}{2000}]{Hamuy2000}
{Hamuy} M.,  {Trager} S.~C.,  {Pinto} P.~A.,  {Phillips} M.~M.,  {Schommer}
  R.~A.,  {Ivanov} V.,   {Suntzeff} N.~B.,  2000, \mn@doi [\aj]
  {10.1086/301527}, \href
  {https://ui.adsabs.harvard.edu/abs/2000AJ....120.1479H} {120, 1479}

\bibitem[\protect\citeauthoryear{{Harris} et~al.,}{{Harris}
  et~al.}{2020}]{Harris2020}
{Harris} C.~R.,  et~al., 2020, \mn@doi [\nat] {10.1038/s41586-020-2649-2},
  \href {https://ui.adsabs.harvard.edu/abs/2020Natur.585..357H} {585, 357}

\bibitem[\protect\citeauthoryear{{Hoogendam} et~al.,}{{Hoogendam}
  et~al.}{2022}]{Hoogendam2022}
{Hoogendam} W.~B.,  et~al., 2022, \mn@doi [\apj] {10.3847/1538-4357/ac54aa},
  \href {https://ui.adsabs.harvard.edu/abs/2022ApJ...928..103H} {928, 103}

\bibitem[\protect\citeauthoryear{{Howell}}{{Howell}}{2001}]{Howell2001}
{Howell} D.~A.,  2001, \mn@doi [\apjl] {10.1086/321702}, \href
  {https://ui.adsabs.harvard.edu/abs/2001ApJ...554L.193H} {554, L193}

\bibitem[\protect\citeauthoryear{{Howell} et~al.,}{{Howell}
  et~al.}{2009}]{Howell2009}
{Howell} D.~A.,  et~al., 2009, \mn@doi [\apj] {10.1088/0004-637X/691/1/661},
  \href {https://ui.adsabs.harvard.edu/abs/2009ApJ...691..661H} {691, 661}

\bibitem[\protect\citeauthoryear{{Hunter}}{{Hunter}}{2007}]{Hunter2007}
{Hunter} J.~D.,  2007, \mn@doi [Computing in Science and Engineering]
  {10.1109/MCSE.2007.55}, \href
  {https://ui.adsabs.harvard.edu/abs/2007CSE.....9...90H} {9, 90}

\bibitem[\protect\citeauthoryear{{Itagaki}}{{Itagaki}}{2020}]{Itagaki2020}
{Itagaki} K.,  2020, Transient Name Server Discovery Report, \href
  {https://ui.adsabs.harvard.edu/abs/2020TNSTR.111....1I} {2020-111, 1}

\bibitem[\protect\citeauthoryear{{Johansson} et~al.,}{{Johansson}
  et~al.}{2013}]{Johansson2013}
{Johansson} J.,  et~al., 2013, \mn@doi [\mnras] {10.1093/mnras/stt1408}, \href
  {https://ui.adsabs.harvard.edu/abs/2013MNRAS.435.1680J} {435, 1680}

\bibitem[\protect\citeauthoryear{{Kelly}, {Hicken}, {Burke}, {Mandel}  \&
  {Kirshner}}{{Kelly} et~al.}{2010}]{Kelly2010}
{Kelly} P.~L.,  {Hicken} M.,  {Burke} D.~L.,  {Mandel} K.~S.,   {Kirshner}
  R.~P.,  2010, \mn@doi [\apj] {10.1088/0004-637X/715/2/743}, \href
  {https://ui.adsabs.harvard.edu/abs/2010ApJ...715..743K} {715, 743}

\bibitem[\protect\citeauthoryear{{Kelsey} et~al.,}{{Kelsey}
  et~al.}{2021}]{Kelsey2021}
{Kelsey} L.,  et~al., 2021, \mn@doi [\mnras] {10.1093/mnras/staa3924}, \href
  {https://ui.adsabs.harvard.edu/abs/2021MNRAS.501.4861K} {501, 4861}

\bibitem[\protect\citeauthoryear{{Kelsey} et~al.,}{{Kelsey}
  et~al.}{2023}]{Kelsey2022}
{Kelsey} L.,  et~al., 2023, \mn@doi [\mnras] {10.1093/mnras/stac3711}, \href
  {https://ui.adsabs.harvard.edu/abs/2023MNRAS.519.3046K} {519, 3046}

\bibitem[\protect\citeauthoryear{{Kim}, {Kang}  \& {Lee}}{{Kim}
  et~al.}{2019}]{Kim2019}
{Kim} Y.-L.,  {Kang} Y.,   {Lee} Y.-W.,  2019, \mn@doi [Journal of Korean
  Astronomical Society] {10.5303/JKAS.2019.52.5.181}, \href
  {https://ui.adsabs.harvard.edu/abs/2019JKAS...52..181K} {52, 181}

\bibitem[\protect\citeauthoryear{{Kochanek} et~al.,}{{Kochanek}
  et~al.}{2017}]{Kochanek2017}
{Kochanek} C.~S.,  et~al., 2017, \mn@doi [\pasp] {10.1088/1538-3873/aa80d9},
  \href {https://ui.adsabs.harvard.edu/abs/2017PASP..129j4502K} {129, 104502}

\bibitem[\protect\citeauthoryear{{Krause}, {Birkmann}, {Usuda}, {Hattori},
  {Goto}, {Rieke}  \& {Misselt}}{{Krause} et~al.}{2008a}]{Krause2008a}
{Krause} O.,  {Birkmann} S.~M.,  {Usuda} T.,  {Hattori} T.,  {Goto} M.,
  {Rieke} G.~H.,   {Misselt} K.~A.,  2008a, \mn@doi [Science]
  {10.1126/science.1155788}, \href
  {https://ui.adsabs.harvard.edu/abs/2008Sci...320.1195K} {320, 1195}

\bibitem[\protect\citeauthoryear{{Krause}, {Tanaka}, {Usuda}, {Hattori},
  {Goto}, {Birkmann}  \& {Nomoto}}{{Krause} et~al.}{2008b}]{Krause2008b}
{Krause} O.,  {Tanaka} M.,  {Usuda} T.,  {Hattori} T.,  {Goto} M.,  {Birkmann}
  S.,   {Nomoto} K.,  2008b, \mn@doi [\nat] {10.1038/nature07608}, \href
  {https://ui.adsabs.harvard.edu/abs/2008Natur.456..617K} {456, 617}

\bibitem[\protect\citeauthoryear{{Lampeitl} et~al.,}{{Lampeitl}
  et~al.}{2010}]{Lampeitl2010}
{Lampeitl} H.,  et~al., 2010, \mn@doi [\apj] {10.1088/0004-637X/722/1/566},
  \href {https://ui.adsabs.harvard.edu/abs/2010ApJ...722..566L} {722, 566}

\bibitem[\protect\citeauthoryear{{Mannucci}, {Della Valle}, {Panagia},
  {Cappellaro}, {Cresci}, {Maiolino}, {Petrosian}  \& {Turatto}}{{Mannucci}
  et~al.}{2005}]{Mannucci2005}
{Mannucci} F.,  {Della Valle} M.,  {Panagia} N.,  {Cappellaro} E.,  {Cresci}
  G.,  {Maiolino} R.,  {Petrosian} A.,   {Turatto} M.,  2005, \mn@doi [\aap]
  {10.1051/0004-6361:20041411}, \href
  {https://ui.adsabs.harvard.edu/abs/2005A&A...433..807M} {433, 807}

\bibitem[\protect\citeauthoryear{McKinney et~al.}{McKinney
  et~al.}{2010}]{McKinney2010}
McKinney W.,  et~al., 2010, in Proceedings of the 9th Python in Science
  Conference. pp 51--56

\bibitem[\protect\citeauthoryear{{Neill} et~al.,}{{Neill}
  et~al.}{2009}]{Neill2009}
{Neill} J.~D.,  et~al., 2009, \mn@doi [\apj] {10.1088/0004-637X/707/2/1449},
  \href {https://ui.adsabs.harvard.edu/abs/2009ApJ...707.1449N} {707, 1449}

\bibitem[\protect\citeauthoryear{{Pan} et~al.,}{{Pan} et~al.}{2014}]{Pan2014}
{Pan} Y.~C.,  et~al., 2014, \mn@doi [\mnras] {10.1093/mnras/stt2287}, \href
  {https://ui.adsabs.harvard.edu/abs/2014MNRAS.438.1391P} {438, 1391}

\bibitem[\protect\citeauthoryear{{Phillips}}{{Phillips}}{1993}]{Phillips1993}
{Phillips} M.~M.,  1993, \mn@doi [\apjl] {10.1086/186970}, \href
  {http://adsabs.harvard.edu/abs/1993ApJ...413L.105P} {413, L105}

\bibitem[\protect\citeauthoryear{{Ponder}, {Wood-Vasey}, {Weyant}, {Barton},
  {Galbany}, {Liu}, {Garnavich}  \& {Matheson}}{{Ponder}
  et~al.}{2021}]{Ponder2020}
{Ponder} K.~A.,  {Wood-Vasey} W.~M.,  {Weyant} A.,  {Barton} N.~T.,  {Galbany}
  L.,  {Liu} S.,  {Garnavich} P.,   {Matheson} T.,  2021, \mn@doi [\apj]
  {10.3847/1538-4357/ac2d99}, \href
  {https://ui.adsabs.harvard.edu/abs/2021ApJ...923..197P} {923, 197}

\bibitem[\protect\citeauthoryear{{Popovic}, {Brout}, {Kessler}, {Scolnic}  \&
  {Lu}}{{Popovic} et~al.}{2021}]{Popovic2021}
{Popovic} B.,  {Brout} D.,  {Kessler} R.,  {Scolnic} D.,   {Lu} L.,  2021,
  \mn@doi [\apj] {10.3847/1538-4357/abf14f}, \href
  {https://ui.adsabs.harvard.edu/abs/2021ApJ...913...49P} {913, 49}

\bibitem[\protect\citeauthoryear{{Price-Whelan} et~al.,}{{Price-Whelan}
  et~al.}{2018}]{astropy:2018}
{Price-Whelan} A.~M.,  et~al., 2018, \mn@doi [\aj] {10.3847/1538-3881/aabc4f},
  \href {https://ui.adsabs.harvard.edu/#abs/2018AJ....156..123T} {156, 123}

\bibitem[\protect\citeauthoryear{{Pskovskii}}{{Pskovskii}}{1977}]{Pskovskii1977}
{Pskovskii} I.~P.,  1977, \sovast, \href
  {https://ui.adsabs.harvard.edu/abs/1977SvA....21..675P} {21, 675}

\bibitem[\protect\citeauthoryear{{Qin} et~al.,}{{Qin} et~al.}{2022}]{Qin2022}
{Qin} Y.-J.,  et~al., 2022, \mn@doi [\apjs] {10.3847/1538-4365/ac2fa1}, \href
  {https://ui.adsabs.harvard.edu/abs/2022ApJS..259...13Q} {259, 13}

\bibitem[\protect\citeauthoryear{{Riess}, {Press}  \& {Kirshner}}{{Riess}
  et~al.}{1995}]{Riess1995}
{Riess} A.~G.,  {Press} W.~H.,   {Kirshner} R.~P.,  1995, \mn@doi [\apjl]
  {10.1086/187704}, \href
  {https://ui.adsabs.harvard.edu/abs/1995ApJ...438L..17R} {438, L17}

\bibitem[\protect\citeauthoryear{{Riess}, {Press}  \& {Kirshner}}{{Riess}
  et~al.}{1996}]{Riess1996}
{Riess} A.~G.,  {Press} W.~H.,   {Kirshner} R.~P.,  1996, \mn@doi [\apj]
  {10.1086/178129}, \href
  {https://ui.adsabs.harvard.edu/abs/1996ApJ...473...88R} {473, 88}

\bibitem[\protect\citeauthoryear{{Rigault} et~al.,}{{Rigault}
  et~al.}{2013}]{Rigault2013}
{Rigault} M.,  et~al., 2013, \mn@doi [\aap] {10.1051/0004-6361/201322104},
  \href {http://adsabs.harvard.edu/abs/2013A%26A...560A..66R} {560, A66}

\bibitem[\protect\citeauthoryear{{Rigault} et~al.,}{{Rigault}
  et~al.}{2015}]{Rigault2015}
{Rigault} M.,  et~al., 2015, \mn@doi [\apj] {10.1088/0004-637X/802/1/20}, \href
  {https://ui.adsabs.harvard.edu/abs/2015ApJ...802...20R} {802, 20}

\bibitem[\protect\citeauthoryear{{Rigault} et~al.,}{{Rigault}
  et~al.}{2020}]{Rigault2020}
{Rigault} M.,  et~al., 2020, \mn@doi [\aap] {10.1051/0004-6361/201730404},
  \href {https://ui.adsabs.harvard.edu/abs/2020A&A...644A.176R} {644, A176}

\bibitem[\protect\citeauthoryear{{Rust}}{{Rust}}{1974}]{Rust1974}
{Rust} B.~W.,  1974, PhD thesis, Oak Ridge National Laboratory, Tennessee

\bibitem[\protect\citeauthoryear{{Schlegel}, {Finkbeiner}  \&
  {Davis}}{{Schlegel} et~al.}{1998}]{SFD1998}
{Schlegel} D.~J.,  {Finkbeiner} D.~P.,   {Davis} M.,  1998, \mn@doi [\apj]
  {10.1086/305772}, \href
  {https://ui.adsabs.harvard.edu/abs/1998ApJ...500..525S} {500, 525}

\bibitem[\protect\citeauthoryear{{Scolnic} et~al.,}{{Scolnic}
  et~al.}{2020}]{Scolnic2020}
{Scolnic} D.,  et~al., 2020, \mn@doi [\apjl] {10.3847/2041-8213/ab8735}, \href
  {https://ui.adsabs.harvard.edu/abs/2020ApJ...896L..13S} {896, L13}

\bibitem[\protect\citeauthoryear{{Scolnic} et~al.,}{{Scolnic}
  et~al.}{2022}]{Scolnic2022}
{Scolnic} D.,  et~al., 2022, \mn@doi [\apj] {10.3847/1538-4357/ac8b7a}, \href
  {https://ui.adsabs.harvard.edu/abs/2022ApJ...938..113S} {938, 113}

\bibitem[\protect\citeauthoryear{{Smith} et~al.,}{{Smith}
  et~al.}{2020}]{Smith2020}
{Smith} M.,  et~al., 2020, \mn@doi [\mnras] {10.1093/mnras/staa946}, \href
  {https://ui.adsabs.harvard.edu/abs/2020MNRAS.494.4426S} {494, 4426}

\bibitem[\protect\citeauthoryear{{Soraisam}, {Matheson}  \& {Lee}}{{Soraisam}
  et~al.}{2021}]{Soraisam2021}
{Soraisam} M.,  {Matheson} T.,   {Lee} C.-H.,  2021, \mn@doi [Research Notes of
  the American Astronomical Society] {10.3847/2515-5172/abf1f7}, \href
  {https://ui.adsabs.harvard.edu/abs/2021RNAAS...5...62S} {5, 62}

\bibitem[\protect\citeauthoryear{{Stritzinger} et~al.,}{{Stritzinger}
  et~al.}{2010}]{Stritzinger2010}
{Stritzinger} M.,  et~al., 2010, \mn@doi [\aj] {10.1088/0004-6256/140/6/2036},
  \href {https://ui.adsabs.harvard.edu/abs/2010AJ....140.2036S} {140, 2036}

\bibitem[\protect\citeauthoryear{{Sullivan} et~al.,}{{Sullivan}
  et~al.}{2006}]{Sullivan2006}
{Sullivan} M.,  et~al., 2006, \mn@doi [\apj] {10.1086/506137}, \href
  {https://ui.adsabs.harvard.edu/abs/2006ApJ...648..868S} {648, 868}

\bibitem[\protect\citeauthoryear{{Sullivan} et~al.,}{{Sullivan}
  et~al.}{2010}]{Sullivan2010}
{Sullivan} M.,  et~al., 2010, \mn@doi [\mnras]
  {10.1111/j.1365-2966.2010.16731.x}, \href
  {https://ui.adsabs.harvard.edu/abs/2010MNRAS.406..782S} {406, 782}

\bibitem[\protect\citeauthoryear{{Tripp}}{{Tripp}}{1998}]{Tripp1998}
{Tripp} R.,  1998, \aap, \href
  {http://adsabs.harvard.edu/abs/1998A%26A...331..815T} {331, 815}

\bibitem[\protect\citeauthoryear{{Uddin}, {Mould}, {Lidman}, {Ruhlmann-Kleider}
   \& {Zhang}}{{Uddin} et~al.}{2017}]{Uddin2017}
{Uddin} S.~A.,  {Mould} J.,  {Lidman} C.,  {Ruhlmann-Kleider} V.,   {Zhang}
  B.~R.,  2017, \mn@doi [\apj] {10.3847/1538-4357/aa8df7}, \href
  {https://ui.adsabs.harvard.edu/abs/2017ApJ...848...56U} {848, 56}

\bibitem[\protect\citeauthoryear{{Uddin} et~al.,}{{Uddin}
  et~al.}{2020}]{Uddin2020}
{Uddin} S.~A.,  et~al., 2020, \mn@doi [\apj] {10.3847/1538-4357/abafb7}, \href
  {https://ui.adsabs.harvard.edu/abs/2020ApJ...901..143U} {901, 143}

\bibitem[\protect\citeauthoryear{{Virtanen} et~al.,}{{Virtanen}
  et~al.}{2020}]{Virtanen2020}
{Virtanen} P.,  et~al., 2020, \mn@doi [Nature Methods]
  {10.1038/s41592-019-0686-2}, \href
  {https://ui.adsabs.harvard.edu/abs/2020NatMe..17..261V} {17, 261}

\bibitem[\protect\citeauthoryear{{Ward} et~al.,}{{Ward}
  et~al.}{2023}]{Ward2022}
{Ward} S.~M.,  et~al., 2023, \mn@doi [\apj] {10.3847/1538-4357/acf7bb}, \href
  {https://ui.adsabs.harvard.edu/abs/2023ApJ...956..111W} {956, 111}

\bibitem[\protect\citeauthoryear{{Wiseman} et~al.,}{{Wiseman}
  et~al.}{2021}]{Wiseman2021}
{Wiseman} P.,  et~al., 2021, \mn@doi [\mnras] {10.1093/mnras/stab1943}, \href
  {https://ui.adsabs.harvard.edu/abs/2021MNRAS.506.3330W} {506, 3330}

\bibitem[\protect\citeauthoryear{{Wolf} et~al.,}{{Wolf}
  et~al.}{2016}]{Wolf2016}
{Wolf} R.~C.,  et~al., 2016, \mn@doi [\apj] {10.3847/0004-637X/821/2/115},
  \href {https://ui.adsabs.harvard.edu/abs/2016ApJ...821..115W} {821, 115}

\bibitem[\protect\citeauthoryear{{Zwicky}}{{Zwicky}}{1939}]{Zwicky1939}
{Zwicky} F.,  1939, \mn@doi [\pasp] {10.1086/124993}, \href
  {https://ui.adsabs.harvard.edu/abs/1939PASP...51...36Z} {51, 36}

\bibitem[\protect\citeauthoryear{{van den Bergh}}{{van den
  Bergh}}{1988}]{vandenBergh1988}
{van den Bergh} S.,  1988, \mn@doi [\apj] {10.1086/166178}, \href
  {https://ui.adsabs.harvard.edu/abs/1988ApJ...327..156V} {327, 156}

\makeatother
\end{thebibliography}

%\begin{thebibliography}{99}
%\end{thebibliography}

%%%%%%%%%%%%%%%%%%%%%%%%%%%%%%%%%%%%%%%%%%%%%%%%%%
%%%%%%%%%%%%%%%%% APPENDICES %%%%%%%%%%%%%%%%%%%%%
\appendix
\section{Further Boot-strapping} \label{further_bootstrap}

\begin{figure*}
    \begin{subfigure}{0.33\textwidth}
        \includegraphics[width=\textwidth]{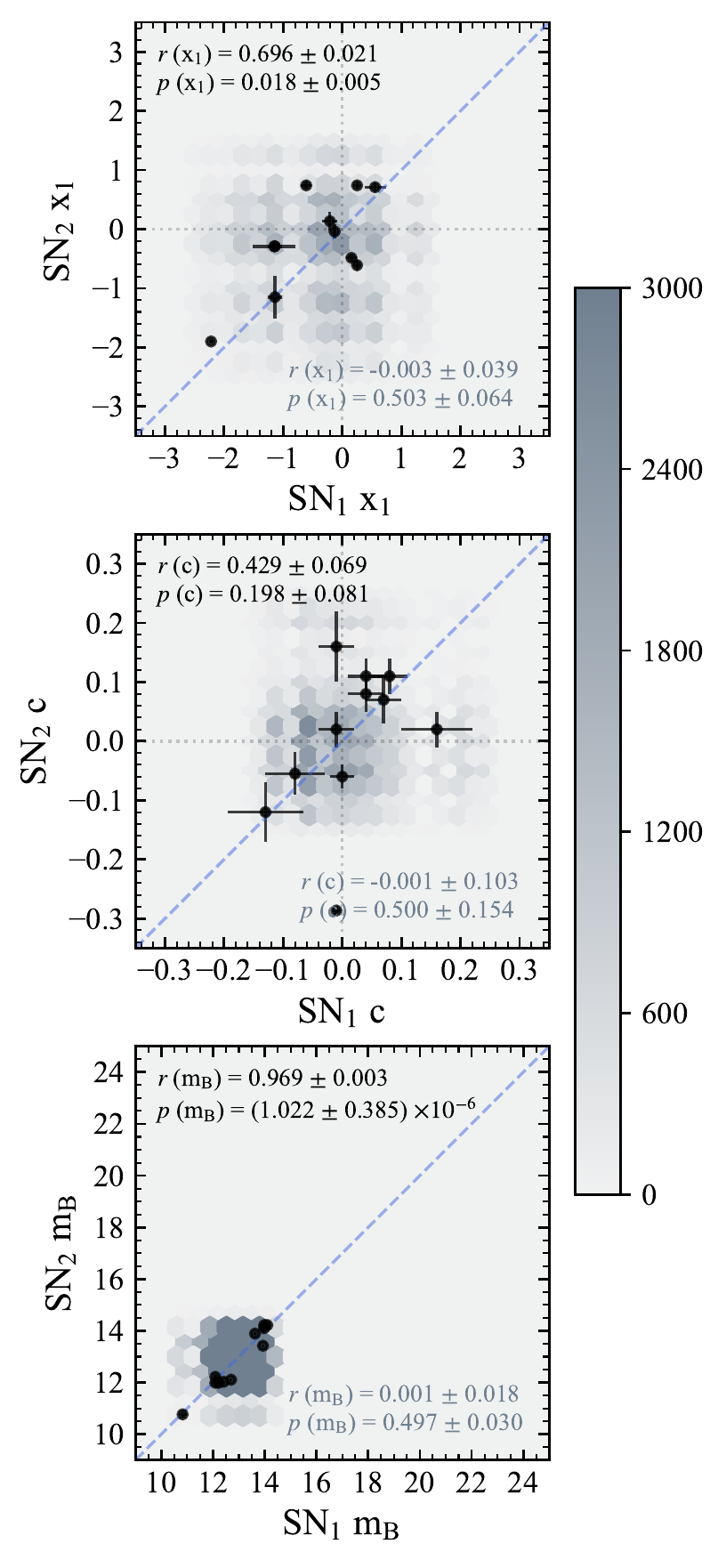}
        \caption{$0 < z \leq 0.01$}
        \label{fig:lowz}
    \end{subfigure}
    \begin{subfigure}{0.33\textwidth}
        \includegraphics[width=\textwidth]{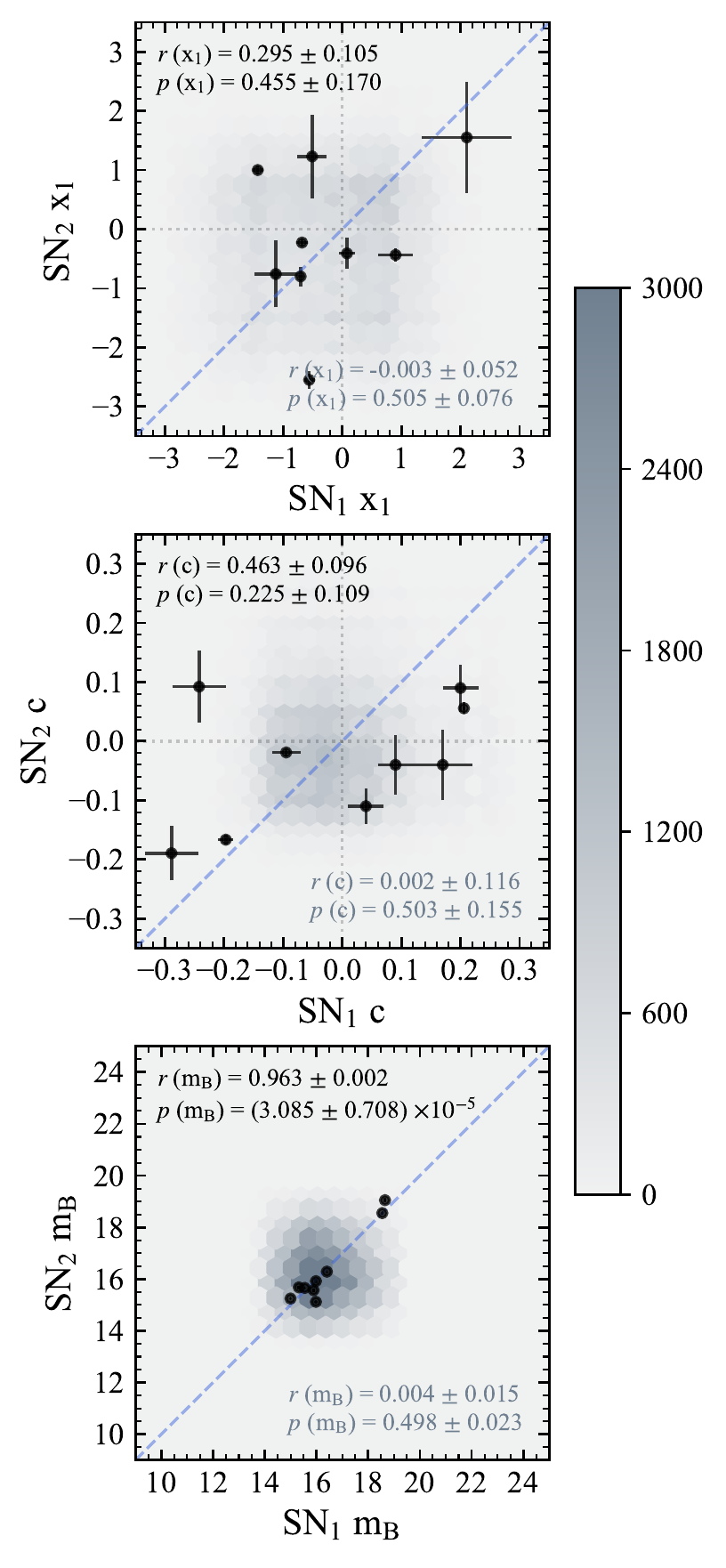}
        \caption{$0.01 < z \leq 0.1$}
        \label{fig:midz}
    \end{subfigure}
    \begin{subfigure}{0.33\textwidth}
        \includegraphics[width=\textwidth]{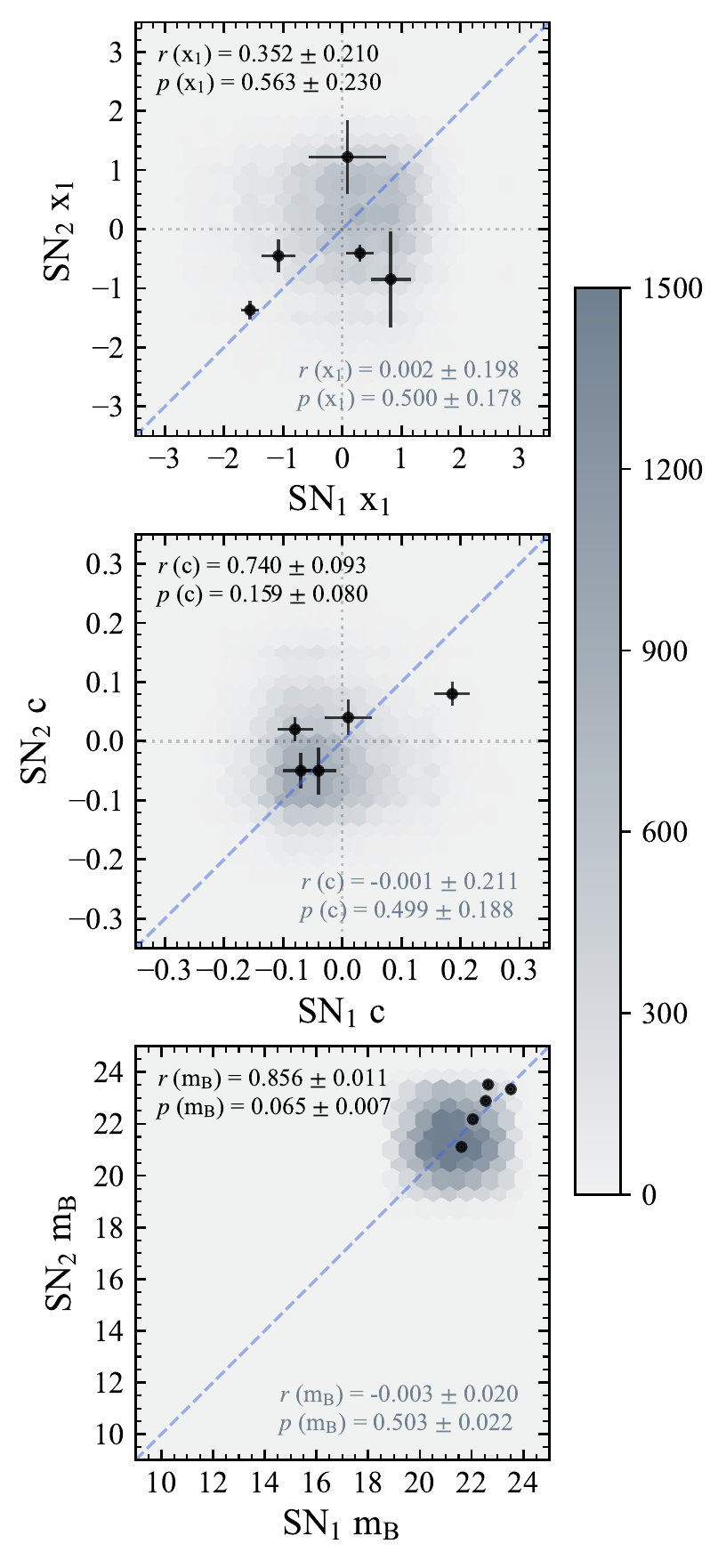}
        \caption{$0.1 < z \leq 0.648$}
        \label{fig:highz}
    \end{subfigure}
\caption{Comparison of light-curve parameters $x_1$ (top panel), $c$ (middle panel) and m$_B$ (lower panel) for all pairs of SN siblings in the cosmology sample, divided into sub-samples by redshift. The blue dashed line represents the $x=y$ line where sibling light-curve parameters are equal. Hexbin data represents the location of the Pantheon+ data in the boot-strapping. Presented in the top left corner for each plot are the Pearson $r$ values of the sibling data, and the same for the boot-strapped Pantheon+ data in the lower right.}
\label{fig:sim-zbins}
\end{figure*}

\comment{As a further test of the boot-strapping analysis discussed in \sref{lc_props}, I break the sample and simulation data into sub-samples of small redshift bins. Given siblings are at the same redshift, whilst the Pantheon+ sample spans a wide redshift range, this test will allow further understanding of the similarities or differences between siblings and random pairs of SNe Ia.

Given the wide redshift range of the siblings, I create three subsamples as follows: a) $0 < z \leq 0.01$, containing 11 pairs, b) $0.01 < z \leq 0.1$, containing 9 pairs, and c) $0.1 < z \leq 0.648$, containing the 5 DES pairs. Breaking down the sample into further sub-samples with a smaller redshift difference creates a large number of bins containing only two pairs of siblings, making statistical comparison difficult. More cosmology-quality siblings will need to be found to improve the statistical power of this type of analysis.

For each defined subsample, I cut down the Pantheon+ sample to match the subsample redshift range, then perform the boot-strapping analysis as outlined in \sref{lc_props} for each case. Comparisons between light-curve parameters in the different subsamples are presented in \fref{fig:sim-zbins}, with the $r$- and $p$-values for the siblings displayed in the top left corners. The Pearson $r$-values and corresponding $p$-values for the sibling data and the boot-strapped data for each bin are presented in \tref{table:z_bin_corrs}, with the $\sigma$ values in \tref{table:z_bins_simulation_comparison}, comparable to \tref{table:simulation_comparison} in the main analysis.}

\begin{table*}
\caption{Pearson $r$-values and corresponding $p$-values for boot-strapping analysis.}
\begin{threeparttable}
\begin{tabular}{cccccc}
\hline
Redshift Bin & Property & \multicolumn{2}{c}{Data} & \multicolumn{2}{c}{Boot-strap}\\
 & & $r$ & $p$ & $r$ & $p$ \\ 
\hline
$0 < z \leq 0.01$  & $x_1$ & $0.696 \pm 0.021$ & $0.018 \pm 0.005$ & $-0.003 \pm 0.039$ & $0.503 \pm 0.064$\\
& $c$ & $0.429 \pm 0.069$ & $0.198 \pm 0.081$ & $-0.001 \pm 0.103$ & $0.500 \pm 0.154$\\
& $\textrm{m}_\textrm{B}$ & $0.969 \pm 0.003$ & $(1.022 \pm 0.385)\times10^{-6}$ & $0.001 \pm 0.018$ & $0.497 \pm 0.030$\\ 
\hline
$0.01 < z \leq 0.1$ & $x_1$ & $0.295 \pm 0.105$ & $0.455 \pm 0.170$ & $-0.003 \pm 0.052$ & $0.505 \pm 0.076$\\
& $c$ & $0.463 \pm 0.096$ & $0.225 \pm 0.109$ & $0.002 \pm 0.116$ & $0.503 \pm 0.155$\\
& $\textrm{m}_\textrm{B}$ & $0.963 \pm 0.002$ & $(3.085 \pm 0.708)\times10^{-5}$ & $0.004 \pm 0.015$ & $0.498 \pm 0.023$\\  
\hline
$0.1 < z \leq 0.648$ & $x_1$ & $0.352 \pm 0.210$ & $0.563 \pm 0.230$ & $0.002 \pm 0.198$ & $0.500 \pm 0.178$\\
& $c$ & $0.740 \pm 0.093$ & $0.159 \pm 0.080$ & $-0.001 \pm 0.211$ & $0.499 \pm 0.188$\\
& $\textrm{m}_\textrm{B}$ & $0.856 \pm 0.011$ & $0.065 \pm 0.007$ & $-0.003 \pm 0.020$ & $0.503 \pm 0.022$\\ 
\hline
\end{tabular}
\end{threeparttable}
\label{table:z_bin_corrs}
\end{table*}

\begin{table*}
\caption{Standard deviation $\sigma$ of the differences in light-curve properties between SN$_\textrm{1}$ and SN$_\textrm{2}$ for the data and for a sample obtained through boot-strapping for a range of different bins divided by redshift.}
\begin{threeparttable}
\makebox[1 \linewidth][c]{       %fixing table spacing
\begin{tabular}{ccccc}
\hline
Redshift Bin & Property &     Data $\sigma$ &              Boot-strap\ $\sigma$ &     Significance of difference\\
\hline
$0 < z \leq 0.01$ & $\Delta x_1$ & $0.392 \pm 0.088$ & $0.711 \pm 0.159$ & $1.759$\\
& $\Delta c$ & $0.081 \pm 0.018$ & $0.073 \pm 0.016$ & $0.337$\\
& $\Delta \textrm{m}_\textrm{B}$ & $0.169 \pm 0.038$ & $0.621 \pm 0.139$ & $3.145$\\
\hline
$0.01 < z \leq 0.1$ & $\Delta x_1$ & $0.792 \pm 0.198$ & $0.842 \pm 0.211$ & $0.174$\\
& $\Delta c$ & $0.083 \pm 0.021$ & $0.072 \pm 0.018$ & $0.392$\\
& $\Delta \textrm{m}_\textrm{B}$ & $0.245 \pm 0.061$ & $0.875 \pm 0.219$ & $2.774$\\
\hline
$0.1 < z \leq 0.648$ & $\Delta x_1$ & $0.500 \pm 0.177$ & $0.614 \pm 0.217$ & $0.408$\\
& $\Delta c$ & $0.041 \pm 0.015$ & $0.057 \pm 0.020$ & $0.639$\\
& $\Delta \textrm{m}_\textrm{B}$ & $0.270 \pm 0.096$ & $0.770 \pm 0.272$ & $1.733$\\
\hline
\end{tabular}
}
\end{threeparttable}
\label{table:z_bins_simulation_comparison}
\end{table*}

\comment{
As can be seen in \fref{fig:sim-zbins}, and highlighted by \tref{table:z_bin_corrs}, the sibling data are more strongly correlated than the boot-strap data for all properties in all redshift bins. This might suggest that the siblings are more similar than random pairs in small redshift bins, however, as in the main analysis, the $p$-values for all properties except $\textrm{m}_\textrm{B}$ are consistent with a lack of correlation in both data and the boot-strap sample. This means that only the $\textrm{m}_\textrm{B}$ values for SN siblings are more similar than any other random pair of SNe, even when only considering subsamples of small redshift bins. I do note however that the $x_1$ $p$-value for the lowest redshift bin may be considered weak evidence that stretch is more correlated for siblings than for random pairs of SNe Ia at low redshift due to being close to the $p$-value threshold when accounting for the uncertainty. %This redshift bin is strongly dominated by Pantheon+ siblings, so may suggest why \citet{Scolnic2022} found that those siblings were $2\sigma$ more similar in $x_1$ than random pairs of SNe Ia in their analysis.

As in the main analysis, I compare between data and boot-strapped standard deviations ($\sigma$) in the differences in light-curve properties between sibling pairs. These are presented in \tref{table:z_bins_simulation_comparison}. For both $\Delta x_1$ and $\Delta c$, the boot-strapped and data $\sigma$ values are consistently less than $2\sigma$ in significance, so it can therefore be concluded that $x_1$ and $c$ values are no more similar for SN siblings than for any random pair of SNe in the same redshift bin. For $\textrm{m}_\textrm{B}$, the data and boot-strapped $\sigma$ values are more consistent in redshift bins than they were when considering the entire sample, with only the lowest redshift bin having $> 3\sigma$ significance indicating siblings being more similar than any random pairs. This bin also contains the siblings with the most similarities in $x_1$ when compared to random pairs at a close to $2\sigma$ difference. These results agree with my findings and conclusions from the main analysis, that sibling SNe Ia are no more similar than any random pair of SNe Ia.
}

%----------
\section{Full Sibling SN Ia Sample} \label{table full sample}

\setlength{\parskip}{0pt}
Here I present the full sibling SN Ia sample, described in \sref{fullsample}. If the set of siblings have previously appeared in the literature as siblings, I give the citation of the associated analyses. Some host galaxies in \comment{DES have not been given names, so I simply refer to these numerically to distinguish between sets of siblings.} Host positions are ra and dec given in degrees. 

\pagebreak
\onecolumn
\clearpage

\begin{longtable}{llp{5cm}p{5cm}}
\caption{Full SN Ia Sibling Sample \label{table:allsibs}}\\
\hline
                Host Galaxy &    Host Position &                          SN Siblings &                      Siblings in literature \\
\hline
\endfirsthead

\hline
\multicolumn{4}{c}{Continuation of \tref{table:allsibs}}\\
\hline
                Host Galaxy &     Host Position &                         SN Siblings &                      Siblings in literature \\
\hline
\endhead

\hline
\multicolumn{4}{c}{\tref{table:allsibs} continued on next page}\\
\hline
\endfoot

\hline
\endlastfoot

         18321526+4347440 &   278.0636 43.79558 &                 SN2019tml, SN2020ksj &                                                    \\
  2MASX J05340703-2936335 &  83.52938 -29.60933 &                 SN2019meo, SN2017ile &                                                    \\
  2MASX J15570268+3725001 &  239.26127 37.41669 &               SN2018cvw, ASASSN-14ay &                                                    \\
              2MFGC 17122 &  341.39196 -8.75072 &                  SN2019smg, SN1995ac &                                                    \\
             A234011+2725 &  355.05004 27.41992 &                    SN2007rx, SN2011L &                                                    \\
                  ARK 098 &   48.10958 19.24606 &                  SN2011ag, SN2020nze &                                                    \\
             CGCG 022-015 &    233.5161 2.21049 &         SN2019vv, SDSS2954-54561-572 &                                                    \\
             CGCG 137-064 &    243.44182 22.919 &      SN2017ets, SN2020kyx, SN2020uoo &                                                    \\
             CGCG 198-046 &  262.82412 36.44289 &                  SN2019iml, SN2009hl &                                                    \\
             CGCG 245-033 &  198.23625 47.45666 &                   PTF11bui, PTF11hub &                                                    \\
             CGCG 266-030 &  152.70707 54.21481 &                 SN2019cdn, SN2018lqj &                                                    \\
             CGCG 291-029 &  166.70645 57.68546 &                  SN2005ea, SN2020yzg &                                                    \\
             CGCG 498-012 &  356.85179 28.39353 &                   SN2004fg, SN2000cw &                                                    \\
                    DES Sibling Host 1 &     40.8636 -1.6024 &               DES13S2dlj, DES14S2pkz &                                \citet{Scolnic2020} \\
                    DES Sibling Host 2 &    53.2948 -27.9576 &               DES14C3zym, DES15C3edd &                                \citet{Scolnic2020} \\
                    DES Sibling Host 3 &    54.4053 -28.3102 &               DES14C2iku, DES17C2jjb &                                \citet{Scolnic2020} \\
                    DES Sibling Host 4 &     41.7614 -1.3781 &               DES15S2okk, DES17S2alm &                                \citet{Scolnic2020} \\
                    DES Sibling Host 5 &    55.1459 -28.6279 &               DES15C2mky, DES16C2cqh &                                \citet{Scolnic2020} \\
                    DES Sibling Host 6 &     6.8201 -42.5739 &                DES13E1wu, DES14E1uti &                                \citet{Scolnic2020} \\
                    DES Sibling Host 7 &    52.2183 -27.5744 &               DES16C3nd0, DES16C3nd1 &                                \citet{Scolnic2020} \\
                    DES Sibling Host 8 &     35.4094 -5.7659 &               DES15X2mlr, DES15X2nku &                                \citet{Scolnic2020} \\
           ESO 119- G 046 &  78.62333 -62.17067 &                  SN2019wcj, SN2011ka &                                                    \\
              ESO 159-G23 &  85.04008 -55.53967 &                ASASSN-15pp, SN2009jz &                                                    \\
              ESO 306-G16 &   84.97267 -40.5125 &                   SN2016yr, SN2013az &                                                    \\
               ESO 427-G6 &  101.4445 -31.23039 &                   SN2004S, SN2020spq &                                                    \\
              ESO 447-G37 & 221.86112 -30.64531 &                   SN2009as, SN2009gb &                                                    \\
               ESO 478-G6 &  32.32533 -23.41508 &                   SN2009le, SN2003em &                                                    \\
               ESO 59-G24 & 120.61504 -72.42492 &          SN2016foy, OGLE-2013-SN-099 &                                                    \\
               ESO 85-G38 &  76.08004 -63.58214 &           SN2006aw, OGLE-2012-SN-051 &                                                    \\
             FAIRALL 0514 & 291.30104 -53.59925 &         SN2015al, PSN J192512-533602 &                                                    \\
                FCCB 1147 &  54.34412 -33.04303 &                    SN2010jd, SN1990Y &                                                    \\
                  IC 1050 &  221.02966 18.01263 &                 SN2021och, SN2020jgs &                                                    \\
                  IC 1222 &  248.78832 46.21391 &                  SN2001dq, SN2020enm &                                                    \\
                  IC 1370 &   318.80946 2.19203 &                  SN2018evw, SN2014bn &                                                    \\
                  IC 1371 &  320.06521 -4.87654 &                  SN2017fms, SN2000ej &                                                    \\
                  IC 1657 &  18.52925 -32.65089 &                  SN2016gfk, SN2012hd &                                                    \\
                  IC 2060 &  64.47233 -56.61625 &                   SN2014dn, SN1997ej &                                                    \\
                  IC 2160 &    88.869 -76.92025 &                    SN2009iw, SN2009J &                                                    \\
                  IC 2597 &  159.44772 -27.0817 &                   AT2015dc, SN2007cv &                                                    \\
                  IC 3900 &  193.92208 27.25076 &                  SN2001cg, SN2020afp &                                                    \\
                  IC 4815 & 286.71087 -61.70136 &                   SN2012fv, SN2011fm &                                                    \\
                   IC 582 &  149.75094 17.81714 &                  SN2019dfa, SN2001gb &                                                    \\
                   IC 988 &   213.63358 3.19024 &                  SN2018qp, SN2016jdv &                                                    \\
             LEDA 1781684 &   59.92542 26.58833 &                  SN2020uk, SN2020rov &                                 \citet{Graham2022} \\
                      M84 &   186.2656 12.88698 &           SN1957B, SN1980I, SN1991bg &                                                    \\
                      M85 &  186.35045 18.19149 &                   SN1960R, SN2020nlb &                                                    \\
           MCG +10-24-007 &  249.28625 58.16783 &                    SN1992R, SN1992ac &                                                    \\
            MCG -02-24-27 &  142.2456 -14.80756 &                  SN2011at, SN2020jgl &                                                    \\
            MCG -02-33-17 & 192.51946 -14.73344 &                  SN2019gbx, SN2008io &                                                    \\
           MCG -03-22-002 & 127.40959 -17.29772 &                  SN2014dx, SN2021kds &                                                    \\
                 NGC 0488 &    20.44521 5.25672 &                    SN2010eb, SN1976G &                                                    \\
                  NGC 105 &    6.31992 12.88386 &                    SN2007A, SN1997cw &                                  \citet{Burns2020} \\
                 NGC 1309 &  50.52733 -15.40006 &                    SN2012Z, SN2002fk &                                                    \\
                 NGC 1316 &  50.67383 -37.20823 & SN2006mr, SN2006dd, SN1980N, SN1981D & \citet{Hamuy1991,Stritzinger2010,Brown2014,Burns2020,Scolnic2022} \\
                 NGC 1404 &  54.71633 -35.59439 &                   SN2007on, SN2011iv &  \citet{Ashall2018,Gall2018,Burns2020,Scolnic2022} \\
                 NGC 1448 &    56.133 -44.64483 &                  SN2001el, SN2021pit &                                \citet{Scolnic2022} \\
                 NGC 1566 &  65.00175 -54.93781 &                 SN2021aefx, SN2010el &                                                    \\
                 NGC 1575 &  66.58562 -10.09844 &                 SN2020sjo, SN2020zhh &                                                    \\
                 NGC 1578 &  65.94437 -51.59944 &                   SN2014cd, SN2013fz &                                                    \\
                 NGC 1954 &  83.20138 -14.06278 &                   SN2010ko, SN2013ex &                        \citet{Brown2014,Burns2020} \\
                 NGC 2076 &  86.69775 -16.78261 &                  SN2017dhr, SN2003hx &                                                    \\
                NGC 2115A &  87.83258 -50.58286 &               ASASSN-15hh, SN2021clw &                                                    \\
                 NGC 2765 &    136.90268 3.3929 &                SN2008hv, ASASSN-13dd &                                  \citet{Brown2014} \\
                 NGC 2841 &  140.51098 50.97652 &                    SN1957A, SN1999by &                                                    \\
                 NGC 3147 &  154.22355 73.40075 &        SN2008fv, SN2021hpr, SN1997bq &                       \citet{Scolnic2022,Ward2022} \\
                 NGC 3190 &  154.52347 21.83229 &                   SN2002bo, SN2002cv &                  \citet{Elias-Rosa2008,Burns2020}  \\
                 NGC 3367 &  161.64564 13.75086 &          SN2018kp, SN2003aa, SN1986A &                                                    \\
                 NGC 3905 &  177.27045 -9.72983 &                    SN2009ds, SN2001E &                                  \citet{Burns2020} \\
                 NGC 3910 &  177.49711 21.33364 &                  SNhunt276, SN2013hl &                                                    \\
                 NGC 3913 &  177.66225 55.35386 &                     SN1979B, SN1963J &                                                    \\
                 NGC 3947 &  178.33467 20.75172 &                     SN2013G, SN2001P &                                                    \\
                 NGC 4076 &   181.1355 20.20495 &                    SN2007M, SN2011bc &                                                    \\
                 NGC 4414 &  186.61292 31.22353 &                     SN2021J, SN1974G &                           \citet{Gallego-Cano2022} \\
                 NGC 4493 &    187.78488 0.6137 &                    SN2004br, SN1994M &                      \citet{Burns2020,Scolnic2022} \\
                 NGC 4636 &   190.70761 2.68778 &                    SN1939A, SN2020ue &                                                    \\
                 NGC 4708 & 192.42283 -11.09303 &                  SN2016cvn, SN2005bo &                                  \citet{Burns2020} \\
                 NGC 4753 &  193.09213 -1.19969 &                     SN1965I, SN1983G &                                                    \\
                 NGC 4898 &  195.07362 27.95539 &                   SN2019be, SN2008dx &                                                    \\
                 NGC 5018 &  198.2543 -19.51819 &       SN2017isq, SN2002dj, SN2021fxy &                         \citet{DerKacy2022}                           \\
                 NGC 5061 & 199.52113 -26.83722 &                    SN2005cn, SN1996X &                                                    \\
                 NGC 5157 &   201.8202 32.03071 &                   SN2020ees, SN1995L &                                                    \\
                  NGC 524 &    21.19883 9.53883 &                    SN2008Q, SN2000cx &                                                    \\
                 NGC 5246 &   204.37236 4.10444 &                  SN2004bk, SN2020dkm &                                                    \\
                 NGC 5253 & 204.98318 -31.64011 &            SN1972E, SN1985B, SN1895B &                                                    \\
                 NGC 5419 & 210.91138 -33.97826 &                  SN2018zz, SN2020alh &                                                    \\
                 NGC 5426 &  210.85354 -6.06911 &                    SN2009mz, SN1991B &                                                    \\
                 NGC 5427 &  210.85854 -6.03081 &                   SN2021pfs, SN1976D &                                                    \\
                 NGC 5442 &   211.1801 -9.71366 &                    SN2011bz, SN2001U &                                                    \\
                 NGC 5468 &  211.64538 -5.45311 &          SN2005P, SN2002cr, SN1999cp &                      \citet{Burns2020,Scolnic2022} \\
                 NGC 5490 &  212.48873 17.54555 &                   SN2015bo, SN1997cn &                    \citet{Burns2020,Hoogendam2022} \\
                 NGC 5557 &  214.60718 36.49356 &                   SN2013gn, SN1996aa &                                                    \\
                 NGC 5643 & 218.16977 -44.17441 &                  SN2017cbv, SN2013aa &                      \citet{Burns2020,Scolnic2022} \\
                   NGC 57 &    3.87862 17.32852 &                   SN2011fp, SN2010dq &                                                    \\
                 NGC 5910 &   229.8531 20.89635 &                  SNhunt175, SN2002ec &                                                    \\
                 NGC 6109 &  244.41882 35.00428 &                   SN2010an, SN2003ia &                                                    \\
                 NGC 6166 &   247.1593 39.55122 &                    PS15aot, SN2009eu &                                                    \\
                 NGC 6240 &   253.24529 2.40093 &                   SN2010gp, PS1-14xw &            \citet{Brown2014,Burns2020,Scolnic2022} \\
                 NGC 6261 &  254.12715 27.97751 &                   SN2008dt, SN2007hu &                                  \citet{Burns2020} \\
                 NGC 6384 &   263.10125 7.06028 &                   SN2017drh, SN1971L &                                                    \\
                 NGC 6521 &  268.95184 62.61225 &                  SN2019cnu, SN2009hs &                                                    \\
                 NGC 6708 &  283.8985 -53.72344 &                   SN2011ce, SN2004do &                                                    \\
                 NGC 6754 & 287.85717 -50.64181 &                   SN2000do, SN1998dq &                                                    \\
                  NGC 692 &   27.17496 -48.6485 &                   SN2007st, SN2010aa &                                                    \\
                 NGC 6956 &  310.97379 12.51192 &      PSN J20435314+1230304, SN2013fa &                                \citet{Scolnic2022} \\
                 NGC 6962 &    311.82943 0.3208 &                   SN2002ha, SN2003dt &                                                    \\
                 NGC 7038 &  318.78129 -47.2205 &                   SN2018hsa, SN1983L &                                                    \\
                 NGC 7311 &   338.52796 5.56989 &                 SN2005kc, SN2020adcb &                                                    \\
                 NGC 7364 &  341.10154 -0.16209 &                   SN2009fk, SN2011im &                                                    \\
                 NGC 7499 &   347.59328 7.58067 &       PSN J23102264+0735202, SN1986M &                                                    \\
                 NGC 7753 &  356.77013 29.48344 &                    SN2006ch, SN2013Q &                                                    \\
                PGC 13297 &    54.00661 1.10476 &                   SN2007jh, SN2010kf &                                                    \\
                PGC 48432 &  205.29641 30.37812 &                    SN2012cf, SN1994O &                                                    \\
                PGC 95451 & 355.91429 -41.32144 &                   SN1997dl, SN1997dk &                                                    \\
 SDSS J023120.85-083613.5 &    37.8369 -8.60378 &                   SN2003kr, SN2002iz &                                                    \\
 SDSS J222012.16+173640.4 &  335.05077 17.61127 &                 SN2019sik, SN2019uej &                                 \citet{Graham2022} \\
                UGC 00881 &    20.22279 4.80134 &                  SN2008gl, SN2020zem &                                                    \\
                UGC 01393 &   28.85282 21.28612 &                  SN2016ijj, SN2003iu &                                                    \\
                UGC 10018 &  235.91479 67.76083 &                   SN2011de, SN2001bs &                                                    \\
                UGC 10166 &  240.96735 39.98444 &                  SN2017po, SN2019pfe &                                                    \\
                UGC 10976 &  266.61613 30.70475 &                  SN2019tdf, SN2013gz &                                                    \\
                UGC 11149 &  272.79071 49.86469 &                   SN2003hs, SN1998dx &                                                    \\
                UGC 12039 &  336.70192 35.51872 &                 SN2019fmm, AT2018gqh &                                                    \\
                UGC 12558 &  350.58033 50.15911 &         SN2007oo, SN2001dt, SN2021re &                                                    \\
                 UGC 1359 &   28.42604 29.93317 &      PSN J01534240+2956107, SN2010hz &                                                    \\
                 UGC 2078 &   38.99954 10.44044 &                 SN2019vmn, SN2019mbm &                                                    \\
                 UGC 2295 &    42.24517 3.16867 &                ASASSN-15mc, SN2003hm &                                                    \\
                 UGC 2828 &   55.60017 39.24453 &                   SN2007kk, SN2006es &                                                    \\
                 UGC 3218 &   75.18212 62.24403 &                    SN2006le, SN2011M &                                  \citet{Burns2020} \\
                 UGC 3329 &   84.13575 16.64886 &                  SN2017ghu, SN1999ek &                                                    \\
                 UGC 3336 &   88.93589 85.91468 &                   SN2006nr, SN2003ij &                                                    \\
                 UGC 3375 &   88.85546 51.91061 &                  SN2016esm, SN2001gc &                                                    \\
                 UGC 3378 &   91.58599 83.83861 &                  SN2010lt, SN2021ltl &                                                    \\
                 UGC 3963 &    115.384 51.68836 &                   SN2008hq, SN2001eo &                                                    \\
                 UGC 4798 &  137.17756 44.81066 &                   SN2013V, SNhunt263 &                                                    \\
                 UGC 5691 &  157.36455 22.00793 &                   iPTF14hvh, SN1991S &                                                    \\
                 UGC 5912 &  162.11371 26.58381 &                    SN2014ea, SN1971U &                                                    \\
                  UGC 595 &   14.39548 -1.39109 &                   SN2007nq, SN2010jo &                                                    \\
                 UGC 6332 &  169.81924 20.81356 &                   SN2007bc, SN2002bp &                                                    \\
                 UGC 6609 &  174.62277 20.52771 &                  SN2018cmk, SN2006bd &                                                    \\
                 UGC 6896 &  178.91679 80.21283 &                   SN2020dic, SN1997T &                                                    \\
                 UGC 7228 &  183.40338 46.49397 &                   SN2007sw, SN2012bh &                      \citet{Burns2020,Scolnic2022} \\
                 UGC 7367 &   184.8837 49.81575 &                 SN2016iuh, SN2019arb &                                                    \\
                 UGC 8204 &   196.91422 6.33595 &                   SN2017hn, PTF12alp &                                                    \\
                 UGC 8670 &  205.42601 40.87447 &                   SN2006dh, SN2003fd &                                                    \\
                 UGC 9396 &    218.9406 24.7258 &         SN2012N, SN2014bg, SN2020mvp &                                                    \\
                 UGC 9625 &   224.43705 6.62693 &                SN2020acai, SN2021jvs &                                                    \\
               VII Zw 737 &  265.42962 67.96196 &                SN2019lcj, SN2020aewj &                                 \citet{Biswas2022} \\
WISEA J011508.03+013006.3 &    18.78293 1.50141 &                 SN2018kly, SN2020cki &                                                    \\
WISEA J021204.64-254440.2 &  33.01921 -25.74453 &                  SN2019mxk, LSQ13bjj &                                                    \\
WISEA J022257.97-041840.7 &    35.7415 -4.31129 &                DES16X3biz, PSc090107 &                                                    \\
WISEA J032542.98+411442.6 &   51.42908 41.24511 &                  SN2019abv, SN2007ry &                                                    \\
WISEA J081757.01+334558.7 &   124.48753 33.7663 &                 SN2020hfs, SN2021abw &                                                    \\
WISEA J123806.26+080245.2 &   189.52572 8.04625 &                 SN2019bbd, SN2020hzk &                                 \citet{Graham2022} \\
WISEA J135842.24+430727.7 &  209.67602 43.12438 &                 SN2020jdq, SN2021meh &                                                    \\
WISEA J153151.48+372445.9 &  232.96422 37.41274 &               AT2017fra, ASASSN-16fc &                                                    \\
WISEA J155106.07+342552.0 &  237.77522 34.43109 &               ASASSN-15ae, SN2016cnv &                                                    \\
WISEA J164445.98-011928.8 &  251.19167 -1.32467 &                 SN2019pab, SN2020nje &                                                    \\
WISEA J191935.19+615048.6 &  289.89643 61.84686 &                 SN2017keg, SN2019tka &                                                    \\

\end{longtable}

%\twocolumn %comment out to get typesetting comment on same page as table
%%%%%%%%%%%%%%%%%%%%%%%%%%%%%%%%%%%%%%%%%%%%%%%%%%
% Don't change these lines
\bsp	% typesetting comment
\label{lastpage}
\end{document}